\documentclass[preprint,12pt]{elsarticle}

\usepackage{color}
\usepackage{lineno}
\usepackage{pdflscape}
\usepackage{amsmath}
\usepackage{booktabs}
\usepackage[hyphens]{url}
\usepackage{hyperref}
\usepackage{color}
\usepackage{multirow}
\usepackage{eurosym}
\usepackage{mathrsfs}

\usepackage{addlines} 
\usepackage{adjustbox}
\usepackage{siunitx}
\usepackage{url}

\usepackage{natbib} 

\usepackage[utf8]{inputenc}
\usepackage[english]{babel}

\usepackage[dvipsnames]{xcolor}
\usepackage{soul}

\usepackage{verbatim}
\definecolor{darkblue}{rgb}{0,0.2,0.4}

\usepackage{longtable}
\usepackage{comment}
\usepackage{placeins}
\usepackage{outlines}
\usepackage{caption}
\usepackage{textcomp,gensymb}
\usepackage{rotating}

\newcommand{\eg}{\textit{e.g.}}
\newcommand{\ie}{\textit{i.e.}}

\newcommand{\ca}{\textit{ca.}}

\usepackage{graphicx}
\usepackage{subcaption}
\usepackage{amssymb}

\journal{Energy (Elsevier)}

\hypersetup{ colorlinks=true,
linkcolor=darkblue,
filecolor=darkblue, 
citecolor=darkblue,
urlcolor=darkblue 
}
\begin{document}

\begin{frontmatter}


\title{Powering Europe with North Sea Offshore Wind: The Impact of Hydrogen Investments on Grid Infrastructure and Power Prices}


\author[a]{Goran Durakovic\corref{cor1}}
\author[a]{Pedro Crespo del Granado}
\author[a]{Asgeir Tomasgard}

\cortext[cor1]{Corresponding author's email address: goran.durakovic@ntnu.no}

\pdfstringdefDisableCommands{%
  \def\corref#1{<#1>}%
}

\address[a]{Dept.~of Industrial Economics and Technology Management, Norwegian University of Science and Technology, Trondheim, Norway}

\begin{abstract}

Hydrogen will be a central cross-sectoral energy carrier in the decarbonization of the European energy system. This paper investigates how a large-scale deployment of green hydrogen production affects the investments in transmission and generation towards 2060, analyzes the North Sea area with the main offshore wind projects, and assesses the development of an offshore energy hub. Results indicate that the hydrogen deployment has a tremendous impact on the grid development in Europe and in the North Sea. Findings indicate that total power generation capacity increases around 50\%. The offshore energy hub acts mainly as a power transmission asset, leads to a reduction in total generation capacity, and is central to unlock the offshore wind potential in the North Sea.
The effect of hydrogen deployment on power prices is multifaceted. In regions where power prices have typically been lower than elsewhere in Europe, it is observed that hydrogen increases the power price considerably. However, as hydrogen flexibility relieves stress in high-demand periods for the grid, power prices decrease in average for some countries. This suggests that while the deployment of green hydrogen will lead to a significant increase in power demand, power prices will not necessarily experience a large increase.

\end{abstract}


\end{frontmatter}



\section{Introduction}

The reduction of greenhouse gas emissions from the energy sector is a central part of the European Commission's strategy to reach climate neutrality by 2050~\cite{EuropeanCommission2018}. To achieve this, renewable energy from wind and solar power will be deployed at an unprecedented scale. In this regard, there are considerable wind resources at sea that offer more consistent production patterns than onshore wind. In fact, the EU has envisioned that 300 GW of offshore wind will be deployed by 2050, where the North Sea region will play a substantial part~\cite{europeancommissionWind2020}.

Wind and solar power plants are intermittent generators subjected to weather conditions (e.g., solar irradiation levels). This presents challenges to the electricity sector, as these conditions do not necessarily coincide with the electricity demand. This implies that either these generators produce power when there is not enough demand, or they may not produce enough in periods of high demand. Hydrogen may help with both of these challenges according to the European Commission, where excess electricity can be used for hydrogen production, storing the energy for periods where renewable generation is insufficient to meet electricity demand~\cite{europeancommission2020}. In this regard, the European Commission has created a hydrogen strategy where they foresee the deployment of 80 GW of electrolyzers by 2030; 40 GW in the EU and 40 GW in neighboring countries who will supply the EU with green hydrogen. This hydrogen can further be used to decarbonize hard-to-abate sectors such as transport and industry reliant on fossil sources as a feedstock. These sectors respectively account for approximately 37.5\% and 22.8\% of total emissions in the EU, further bridging the gap to the European Commission's plan of climate neutrality in the EU by 2050.

The European Commission's planned decarbonization of the power system and simultaneous build-up of electrolyzer capacity raises questions about how the power grid will develop, and what the effects on the power prices will be. To this end, this paper studies the high renewable deployment in combination with a demand for green hydrogen towards 2060. The offshore energy hub and wind-hydrogen synergies may be central to maximize the value of North Sea offshore wind, and so the impact of the energy hub on the wind-hydrogen investments and other grid infrastructure is considered. In other words, the paper focuses on the following research questions:

\begin{itemize}
    \item How are investments into European power generation and transmission affected by the deployment of green hydrogen generation?
    \item How does the North Sea region contribute to the European strategies on hydrogen production and power sector decarbonization?
    \item To what extent can green hydrogen production lower curtailment\footnote{In this work, curtailment is defined as the unused available power from intermittent renewable sources.} of offshore wind farms in the North Sea?
    Can the offshore energy hub also be used to reduce curtailment of offshore wind farms?
\end{itemize}

These questions are addressed using a capacity expansion model that closely links the power sector with green hydrogen production. The model co-optimizes the development of both the European power network and hydrogen supply chains, subject to the European decarbonization goals for the power sector. This provides the framework to investigate the role of the North Sea region in providing low-carbon energy to Europe under the demand projections for hydrogen towards 2060.

\section{Literature review}



Techno-economic studies into hydrogen production are widespread in the literature. These studies tend to analyse the cost of hydrogen in a given system under different individual scenarios, or minimize the levelized cost of hydrogen for a given production plant \eg~\cite{Yoshida2022} that analyzed the operations of an electrolyzer in Japan or \cite{Karayel2022} that investigated using geothermal energy for hydrogen production. Other papers study offshore production of hydrogen, such as~\cite{meier2014} who studied the production of hydrogen on an offshore platform in Norway, \cite{babarit2018} who studied the production of hydrogen on autonomous offshore vessel and \cite{woznicki2020} who studied the placement of electrolyzers and batteries. Some have focused on the best way to transmit the offshore wind energy, where~\cite{yan2021} compared export via a cable, or in the cases with hydrogen production, where the hydrogen should be produced and how it should be transported. Similarly,~\cite{damore2021} studied the ranges under which cable electricity transport, hydrogen ship transport and hydrogen pipeline transport are competitive against each other. \cite{singlitico2021} compared the system costs for green hydrogen production when the electrolyzer was placed onshore, offshore on an energy hub and in-turbine. A common finding in these studies is that green hydrogen production from offshore wind systems is an efficient way to transmit the wind energy, and that hydrogen can be economically produced on a centralized offshore hub and exported via ships and pipelines. This is in line with the review by~\cite{Calado2021b}.

Optimization techniques have been applied to hydrogen infrastructure problems. Some studies focus on certain parts of the supply chain and model that in high detail, \eg \ optimizing the hydrogen transmission or distribution networks such as~\cite{andre2013} who developed an approach for designing a hydrogen pipeline network, and~\cite{baufume2013} who used designed a hydrogen pipeline network in Germany using GIS. Furthermore,~\cite{reuss2019} compared the modelling of hydrogen pipelines with linear and non-linear approaches, finding that linear approaches are largely appropriate, but that their solutions should be checked for feasibility. Other papers have focused on optimizing the entire supply chain, including production, storage and transport. Of these, some are deterministic and only feature a snapshot view of the system (\ie, no temporal dynamics), such as~\cite{almansoori2006}, who published among the first papers on hydrogen infrastructure optimization, focusing on the United Kingdom, \cite{kim2008b} who used a similar approach on the South Korean market, \cite{dayhim2014} who introduced uncertain demand but kept the snapshot view and \cite{kim2017} who expanded the South Korean model. Subsequent extensions to these works include reformulations into multi-stage optimization problems, such as~\cite{Almansoori2009} who extended their initial model to include several stages and \cite{kim2016} who also similarly extended their South Korean model. Later, these models were expanded to include stochastic demand as well, such as \cite{nunes2015} that introduced this in the UK model and \cite{kim2008a} that implemented the same in the South Korean model. It is common in this literature that electricity is assumed available at a fixed price, but new studies have started to couple the power and green hydrogen sectors explicitly in their optimization models \eg~\cite{tlili2019} that investigated the role of interconnections on how much excess electricity would be available for hydrogen production in France, ~\cite{pan2020} that developed a model that co-optimizes investments into the electricity and hydrogen systems and \cite{xiong2021} that studied the use of power-to-gas to lower curtailment of renewable generators in Germany and transport hydrogen across congested electricity lines.

There is also an important literature stream focusing on optimization in hydrogen supply chains, see comprehensive reviews in~\cite{agnolucci2013}, \cite{Li2019}, and \cite{FODSTAD2022112246}. A recurrent challenge noted in these reviews  is that uncertainty (\eg \ uncertainty in renewable generation or hydrogen production) is disregarded in optimization models. Furthermore, studies usually focus on the domestic demand and production of a single country, neglecting the possibility of international trade of electricity and hydrogen. Few works also consider the temporal evolution of the system, and instead only optimize for a snapshot. The review authors argue that the system's evolution over time should be explicitly modelled. Altogether, the review authors recommend that future works considers both short-term operational uncertainty and multi-stage investments.

Based on the reviews by~\cite{agnolucci2013} and~\cite{Li2019}, and this literature review, there are important research gaps to highlight. Hydrogen demand is expected to grow in the future, and it is important to plan the infrastructure deployment coherently over time. This requires a multi-stage model, which allows for the re-use of previously built infrastructure. Furthermore, while techno-economic analyses can highlight potentially viable strategies of hydrogen production, it is difficult to compare different techno-economic analysis to get an overview of best hydrogen deployment options. It is therefore suggested to use optimization mechanisms in which the main ways of producing and transporting hydrogen can be compared endogenously and consistently. Since hydrogen production from electrolyzers will undoubtedly closely interlink with the electricity grid, it is important to model these together, rather than having one be an exogenous input to the other \cite{FODSTAD2022112246}. Also, when modelling the electricity grid, the stochastic nature of renewable generators should not be ignored, as there will be times when these do not generate power. Finally, international trade of both electricity and hydrogen may significantly reduce overall system costs, which is not addressed in the literature.

To address these research gaps, hydrogen production and demand are implemented in a capacity expansion model for the European power sector. This implementation will feature hydrogen production to satisfy a hydrogen demand, but also the use of hydrogen in the power sector where economical. The model represents the stochasticity associated with renewable power generation in a detailed way, and features the simultaneous build-up of both electric and hydrogen infrastructure until 2060. Table \ref{tab:literaturecomparison} compares this work with the most relevant available literature and summarizes key differences. In short, this paper main contributions are as follows:
\begin{itemize}
    \item Closely linking hydrogen production with the electricity market at the Pan-European level.
    \item Analyze the increase in European demand for hydrogen and power, and associated generation and transmission capacity to meet these demands.
    \item Incorporate strategic and operational aspects in decision-making.
    \item Include a detailed representation of the intermittency and stochasticity of renewable generators.
    \item Allow for trade of electricity and hydrogen within Europe.
    \item Provide new insights into the role of the North Sea region as a power source for green hydrogen.
\end{itemize}

To summarize, this paper fills several of the research gaps highlighted by~\cite{agnolucci2013} and \cite{Li2019}, including \eg \ considering short-term operational uncertainty, while simultaneously including the long-term multi-period perspective on investments. This approach more closely aligns with the decisions that relevant market actors will make, and so this work provides new insights into the cost-optimal evolution of the hydrogen supply chain.

\newpage
\begin{sidewaystable}[ht!]
    \centering
    \begin{tabular}{llccccc}
    \hline
        \multirow{2}{*}{\textbf{Ref.}} & \multirow{2}{*}{\textbf{Authors}} & \multirow{2}{*}{\textbf{Optimization\footnote{
        Here, those works that use linear programming, mixed-integer linear programming or other non-linear optimization techniques different from individual case analysis are considered as optimization.}
        }} &   \multirow{2}{*}{\textbf{Multi-period\footnote{Multi-period is defined as including a growth in hydrogen demand, and showing how hydrogen generation and transmission changes to accommodate this increase.}
        }} & \multirow{2}{*}{\textbf{Stochastic}} & \textbf{Integrated}  & \textbf{International} \\
        & & & & & \textbf{el. grid} & \textbf{trade} \\
        \cite{almansoori2006,Almansoori2009,almansoori2012} & Almansoori \& Shah & X & X & X & & \\
        \cite{kim2016,kim2017} & Kim et al. & X & & & & \\
        \cite{tlili2019} & Tlili et al. & X & & & X & X \\
        \cite{pan2020} & Pan et al. & X & & X & X & \\
        \cite{weimann2021} & Weimann et al. & X & & & X & \\
        \cite{greiner2007} & Greiner et al. & & & & X & \\
        \cite{meier2014} & Meier & & & & X & \\
        \cite{singlitico2021} & Singlitico et al. & & & & X & \\
        
        \multicolumn{2}{c}{This paper} & X & X & X & X & X \\
        \hline
    \end{tabular}
    \caption{Comparison of this paper with most relevant literature.}
    \label{tab:literaturecomparison}
\end{sidewaystable}

\FloatBarrier
\newpage
\section{Modelling framework} 

A multi-horizon capacity expansion model is implemented to investigate both the investments and operations of power and hydrogen infrastructure. This model is explained in detail in this section.


\subsection{The EMPIRE capacity expansion model}
\label{sec:EMPIRE}

This work uses the EMPIRE model~\cite{Backe2022, Skar2016b, Skar2014}, a tool that allows for the analysis of lowest-cost energy transitions in Europe. The entire code along with all data parameters can be found in~\cite{projectGithub}. EMPIRE is a stochastic~\cite{Birge2011} multi-horizon program~\cite{Kaut2014b} that minimizes investment costs and operating costs from power generators, power transmission lines and energy storage units. Multi-horizon stochastic programs allow for uncertainty in both long term, strategic periods as well as the short term operational time steps. 

EMPIRE solves the hourly dispatch for representative weeks of the year, \ie \ one week for winter, spring, summer and autumn. Additionally, to ensure that the investment decisions are valid even for high-demand hours that may otherwise not be included in the representative seasonal week, EMPIRE also includes two peak days (two 24 hours days) that feature peak load demand profiles. These representative days are scaled up so that the operational costs account for one representative year of operation. There is one such representative year per stochastic scenario. These operational years are further embedded into longer investment periods that typically last for five years in EMPIRE. 

The uncertain elements in the EMPIRE model are assumed to be complex and mutually dependent. In order to preserve the autocorrelation for these stochastic processes, the stochastic sampling is done chronologically. This means that the sampling starts by choosing a random hour for each season, and then including the following hours in the sample, so that the length of the sample matches the length of the season (\ie, 168 hours for the regular seasons, and 24 hours for the peak seasons). The same hours are sampled for each node to preserve the spatial correlation. Furthermore, the same hours are also sampled for each of the stochastic processes, in order to preserve the correlation between these.

\subsection{Hydrogen module in EMPIRE}
\label{sec:hydrogen}

This subsection gives an overview of the modifications made to the EMPIRE code to include the hydrogen economy. Section \ref{sec:constraints} details the implementation of additional constraints and modifications to existing ones, while Section \ref{sec:objectivefunction} explains the changes made to the objective function.

\subsubsection{Additional constraints}
\label{sec:constraints}

Each node has an exogenous annual hydrogen demand, which must be met by the local hydrogen production, as shown in Equation \ref{eqn:h2demand}: 

\begin{equation}
    \label{eqn:h2demand}
    \sum_{s \in \mathscr{S}} \alpha_s \sum_{h \in \mathscr{H}} y^{hyd,sold}_{n,h,i,\omega} \geq D^{hyd}_{n,i} 
    \quad 
    \forall n \in \mathscr{N}^{hyd}, i \in \mathscr{I}, \omega \in \Omega
\end{equation}

where $ \alpha_s$ is the seasonal scaling factor, $y^{hyd,sold}_{n,h,i,\omega}$ is the amount of hydrogen sold to satisfy the exogenous demand in node $n$ in hour $h$ in period $i$ in scenario $\omega$, and $D^{hyd}_{n,i}$ is the annual hydrogen demand in node $n$ in period $i$.

Note that there is an implied flexibility in the above formulation, arising from the mismatched timescales between the hourly hydrogen production and the annual hydrogen demand. The model is thus more flexible than if the hydrogen demand was given hourly. This flexibility much be priced, and this is shown later in Equation \ref{eqn:operationalstoragecost} in Section \ref{sec:objectivefunction}. Demand is given annually because the data source~\cite{Wang2021} only offered this resolution. Hydrogen is exclusively produced through electrolysis. Hydrogen production is linked to the power market with the following constraint:

\begin{equation}
    y^{hyd}_{n,h,i,\omega} = \frac{y^{p2h}_{n,h,i,\omega}}{\eta^{p2h}_i}
    \quad
    \forall n \in \mathscr{N}^{hyd}, h \in \mathscr{H}, i \in \mathscr{I}, \omega \in \Omega
    \label{eqn:powerlinking}
\end{equation}

where $y^{hyd}_{n,h,i,\omega}$ is the amount of hydrogen produced in node $n$ in hour $h$ in period $i$ in scenario $\omega$,  $y^{p2h}_{n,h,i,\omega}$ is the amount of power used for hydrogen production in node $n$ in hour $h$ in period $i$ in scenario $\omega$, and $\eta^{p2h}_i$ is the efficiency of the electrolyzer in period $i$.

Equation \ref{eqn:powerlinking} links hydrogen production to the power sector in EMPIRE. In each operational hour, the model optimizes the generation of dispatchable generators, and chooses whether the power generated should be used in the power sector to satisfy power demand, or whether it should be used for hydrogen production. This choice is especially pertinent for those instances where there is a large share of renewable generation, as for these generators there is the third option of curtailing generation.

It is also necessary to invest in electrolyzer capacity in order to use power to produce hydrogen, and this is ensured by the following equation:

\begin{equation}
    y^{p2h}_{n,h,i,\omega} \leq v^{el}_{n,i}
    \quad
    \forall n \in \mathscr{N}^{hyd}, h \in \mathscr{H}, i \in \mathscr{I}, \omega \in \Omega
\end{equation}

where $v^{el}_{n,i}$ is the available electrolyzer capacity in node $n$ in period $i$.

Note that in EMPIRE, the installed capacity of an asset $a$ in period $i$ is generally defined as follows:

\begin{equation}
    v^a_{n,i} = \bar{x}^{a}_{n,i} + \sum_{j=i'}^i x^a_{n,i}
    \quad
    \forall n \in \mathscr{N}, i \in \mathscr{I}, i' \in \max\{1,i-i^{life}_a\}
\end{equation}

where $\bar{x}^{a}_{n,i}$ is the remaining available capacity from the initial capacity, which decreases according to the lifetime of the asset, and $x^a_{n,i}$ is the amount of capacity invested into in node $n$ in period $i$. 

The hydrogen market is also subjected to flow constraints, \ie \ hydrogen cannot appear from nothing, but must be either produced or transported in order for it to be consumed in a node. This is implemented like the equation below, which includes hydrogen production, hydrogen selling, hydrogen consumption for power production and hydrogen transport:

\begin{multline}
    y^{hyd}_{n,h,i,\omega} - y^{hyd,sold}_{n,h,i,\omega} 
    - y^{hyd,h2p}_{n,h,i,\omega} - \sum_{n_2 \in \mathscr{A}^{hyd}_n}  y^{hyd,trans}_{n,n_2,h,i,\omega} - y^{hyd,trans}_{n_2,n,h,i,\omega} = 0 \\
    \forall n \in \mathscr{N}^{hyd}, h \in \mathscr{H}, i \in \mathscr{I}, \omega \in \Omega 
    \label{eqn:h2flowbalance}
\end{multline}

where $y^{hyd,trans}_{n,n_2,h,i,\omega}$ is the amount of hydrogen transmitted from node $n$ to node $n2$ in hour $h$ in period $i$ in scenario $\omega$.

Equation \ref{eqn:h2flowbalance} also includes the variable $y^{hyd,h2p}_{n,h,i,\omega}$, which describes the use of hydrogen to produce power. This hydrogen does not have any CO$_2$ emissions, and so is a new emission-free fuel in EMPIRE's power sector. Previous research has shown that using hydrogen from electrolyzers in the power system is typically not economical, but this option is included in this model in case hydrogen is needed for flexibility reasons. The cost parameters are the same as for regular gas turbines, which can be seen as an optimistic scenario.

Equation \ref{eqn:h2flowbalance} includes hydrogen transport. These variables are bound by the capacity that has been invested into the (bidirectional) pipeline:

\begin{equation}
    y^{hyd,trans}_{n,n_2,h,i,\omega} \leq v^{hyd,trans}_{n,n_2,i} \vee v^{hyd,trans}_{n_2,n,i}
    \quad
    \forall (n,n_2) \in \mathscr{L}^{hyd}, h \in \mathscr{H}, i \in \mathscr{I}, \omega \in \Omega
\end{equation}

Compressors are needed in order to transport hydrogen through pipelines over large distances, and these compressors require power to function. This power load is shared equally between the sending and receiving nodes, and is part of the electric energy balance for these two nodes.

\subsubsection{Additions to the objective function}
\label{sec:objectivefunction}

Three additions are made to the objective function of EMPIRE to include the hydrogen market. The first addition is the investment costs of the electrolyzers and pipelines. The model finds that optimal timing and capacity for these, and the discounted costs of these are added to the objective function. This contribution is shown in Equation \ref{eqn:operationaldiscount}. 

\begin{equation}
    \label{eqn:investmentcost}
    \sum_{i\in \mathscr{I}} \frac{1}{(1+r)^{(i-1)*L^{period}}} \times \left(\sum_{n\in \mathscr{N}^{hyd}}C^{el}_{i} \times x^{el}_{n,i} + \sum_{l\in \mathscr{L}^{hyd}} C^{pipe}_{l,i} \times x^{pipe}_{l,i}\right)
\end{equation}

Here $r$ is the discount factor, $L^{period}$ is the length of each period, $\mathscr{N}^{hyd}$ is the set of nodes that can produce hydrogen, $\mathscr{L}^{hyd}$ is the set of all hydrogen links, $C^{el}_{i}$ is the linearized cost of electrolyzers in period $i$, $x^{el}_{n,i}$ is the electrolyzer capacity invested into in node $n$ in period $i$, $C^{pipe}_{l,i}$ is the linearized cost of a hydrogen pipeline for link $l$ in period $i$ and $x^{pipe}_{l,i}$ is the pipeline capacity invested into for link $l$ in period $i$.

The costs $C^{el}_{i}$ and $C^{pipe}_{l,i}$ include all the fixed operations and maintenance (O\&M) costs for the assets, and these have been properly discounted for the lifetimes of the assets using the discount factor in the model. Furthermore, in this work, only proton exchange membrane (PEM) electrolyzers are considered. For these electrolyzers, the lifetime of the plant is significantly longer than the lifetime of the electrolyzer stacks, and so the latter must be replaced during the electrolyzer's lifetime~\cite{Buttler2018}.  Note that hydrogen distribution (e.g., pipes) might sometimes have leakage situations and related costs. EMPIRE does not consider this detail modelling; we assume that this is minimum and security aspects of hydrogen technology is mature.


The next component added to the objective function is operational. Since EMPIRE groups several years into periods where the operations within each period are identical (\eg \  2025-2030), any costs or revenues within the periods need to be discounted to the start of that period. This is done using an operational discount factor, $\varphi$, which is defined in Equation \ref{eqn:operationaldiscount}. This factor is multiplied with any costs or revenues.

\begin{equation}
    \label{eqn:operationaldiscount}
    \varphi = \sum_{j=0}^{L^{period}-1} \frac{1}{(1+r)^j}    
\end{equation}

The flexibility that is  present in Equation \ref{eqn:h2demand} implies the use of a hydrogen storage to meet the demand. These storages are not explicitly modelled in this work, because they would not be used with this implementation of hydrogen demand. Instead, the costs of this storage is included in the objective function by charging an average cost for storage ($\Bar{C}^{storage}$) to each unit of hydrogen produced ($y^{hyd}_{n,h,i,\omega}$). Because this is a stochastic program, the expectation of this cost is included in the objective function:

\begin{equation}
    \label{eqn:operationalstoragecost}
    \sum_{i\in \mathscr{I}} \frac{1}{(1+r)^{(i-1)*L^{period}}} \times \sum_{\omega \in \Omega}\sum_{s \in \mathscr{S}}\sum_{h \in \mathscr{H}}\sum_{n\in \mathscr{N}^{hyd}} \pi_\omega \times\varphi \times \alpha_s \times  \Bar{C}^{storage}*y^{hyd}_{n,h,i,\omega}
\end{equation}

where $\pi_\omega$ is the probability of scenario $\omega$.

As EMPIRE only models a subset of the hours in a year, it is necessary to scale up the operational costs and revenues so that they represent the whole year. $\alpha_s$ is this scaling factor for each season. 

The hydrogen module in EMPIRE is demand-driven, meaning that the optimal investment decisions are only made to satisfy the exogenous demand, as shown in Equation \ref{eqn:h2demand}.

\subsection{Offshore energy hub}
\label{sec:offshore_energy_hub}

Energy hubs are modelled with linear variables and cost parameters; no integer variables are used. Building an offshore energy hub will in reality result in a large fixed cost. This is not included for computational reasons. Instead, the cost of the energy hub is reflected through the cost of the electrical converter located at the hub, as shown in equation \ref{eqn:hubcost}: 

\begin{equation}
    \label{eqn:hubcost}
    \sum_{i \in \mathscr{I}} \frac{1}{(1+r)^{(i-1)*L^{period}}} \times C^{conv}_{i} \times x^{conv}_{i}
    \quad
    \forall i\in \mathscr{I}
\end{equation}

where $C^{conv}_{i}$ is the linearized cost of the offshore converter in period $i$, and $x^{conv}_{i}$ is the amount of converter capacity bought in period $i$.

Furthermore, Equation \ref{eqn:convertercapacity} ensures that the electrical infrastructure on the hub is sufficiently dimensioned for the power transmission and the electrolyzer capacities. Most likely the converter will be oversized with this constraint -- it is unlikely that the the cables connected to the hub will be fully utilized while the electrolyzer is used at full capacity. However, by over-dimensioning the converter, the cost will be closer to the large fixed cost that would be required to build the hub.

\begin{equation}
    \label{eqn:convertercapacity}
    x^{el}_{n,i} + \sum_{n_2 \in \mathscr{L}_n} x^{trans}_{n,n_2,i} + x^{trans}_{n_2,n,i} \leq x^{conv}_{i}
    \quad
    \forall n\in \mathscr{N}^{hub}, i \in \mathscr{I}
\end{equation}

\section{Case study and data}
\label{sec:casestudy}

The focus is to study the option to build an offshore energy hub in the North Sea, and the effect of a European demand for green hydrogen. This defines four cases, where the difference between each case is whether the offshore energy hub and hydrogen are included or not. When the hydrogen demand is included, it is assumed that all hydrogen is produced through electrolysis and there are no hydrogen imports.

%

\subsection{EMPIRE model implementation}
The model considers eight investment periods, each lasting for five years from 2020 to 2060. A discount rate of 5\% is assumed, following~\cite{Backe2022,Skar2014}. The model features 50 nodes; 30 nodes for all countries in the EU-27 excluding Cyprus and Montenegro, but with Bosnia-Herzegovina, Great Britain, North Macedonia, Serbia, Switzerland included; five nodes for the Norwegian price zones in Nord Pool; 14 nodes for the North Sea offshore wind farms; and one node for the offshore energy hub. A CO$_2$ cap is implemented, starting at 1110 Mton CO$_2$eq. per year and ending at 22 Mton CO$_2$eq. per year, in line with~\cite{EuropeanCommission2018}. Power demand is flexible, but any demand not met is penalized with a cost of \euro22,000/MWh, following~\cite{LondonEconomics2013}.

\subsection{Hydrogen data and assumptions}
The hydrogen demand is an exogenous parameter. Forecasting the future hydrogen demand is challenging and outside the scope of this work. Therefore, existing literature on future hydrogen demand is used, which includes industrial and transport hydrogen demands in the European countries~\cite{Wang2021}.

This demand is met by producing hydrogen using electrolysis. There is no explicit constraint that ensures that all hydrogen is produced from green energy. Instead, the modelled power market is subjected to the European Commission's vision for a decarbonized power system~\cite{EuropeanCommission2018}.

Wind power production is highly variable, and any electrolyzer system that relies on wind as its main power source will need to be flexible enough to follow the wind power generation. PEM electrolyzers are much more flexible than alkaline electrolyzers~\cite{Bertuccioli2014} and solid oxide electrolyzers~\cite{Buttler2018} in terms of both ramping rates and start-up times. This makes PEM electrolyzers particularly suitable for wind powered electrolysis out of these three technologies, and therefore only PEM electrolyzers are considered in this work. 

After production, hydrogen is transported via pipelines. The compressors that ensure the flow through the pipelines require power, and this power load is shared equally between the sending and receiving nodes.


\subsection{Modelling the North Sea}
\label{sec:north_sea}

\begin{figure}
    \centering
    \includegraphics[width=.9\textwidth]{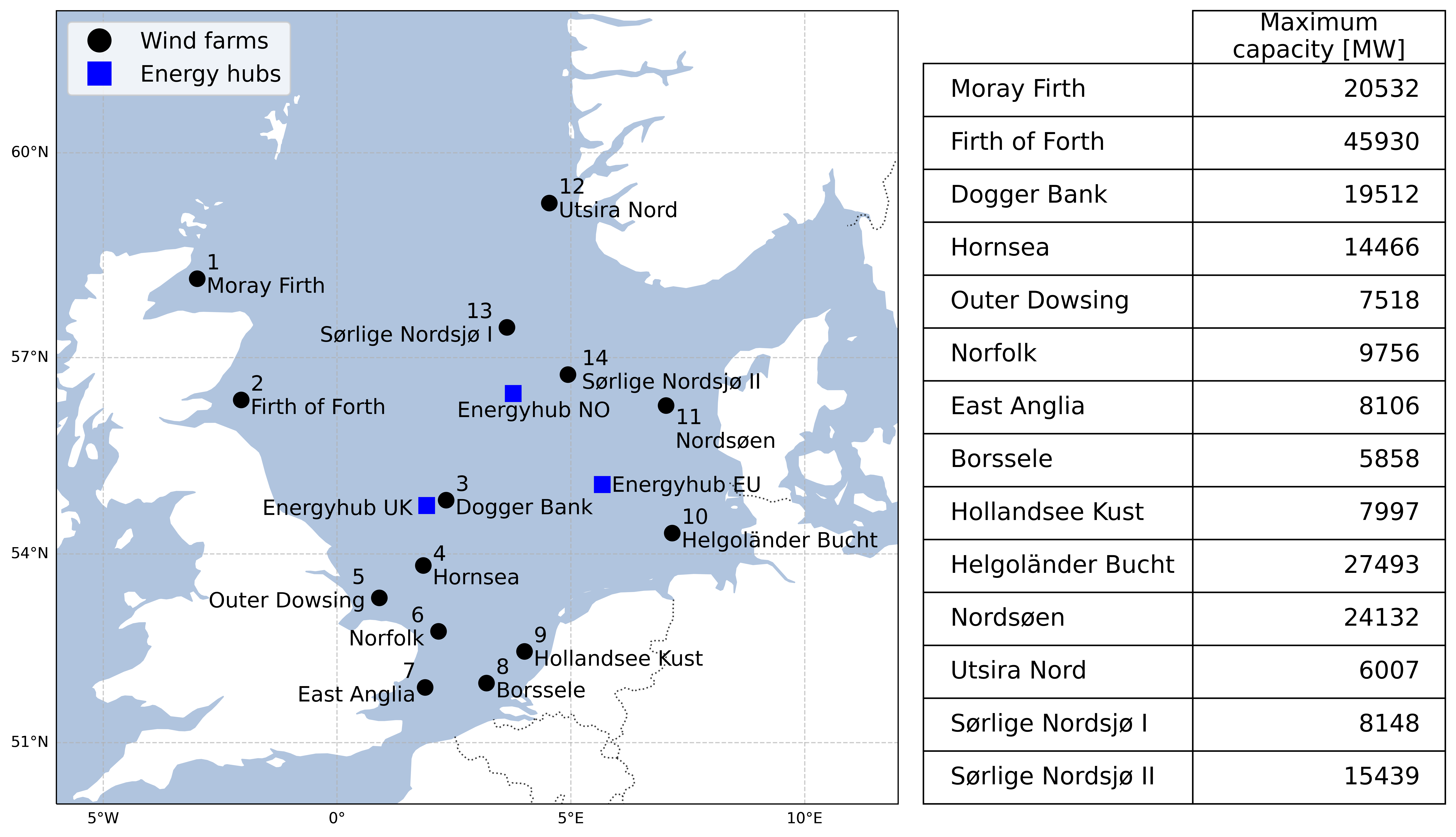}
    \caption{Location of included wind farm sites and considered energy hubs.}
    \label{fig:windfarms}
\end{figure}

A detailed representation of the North Sea offshore wind farms are represented in this paper, see Figure \ref{fig:windfarms}. They are modelled as individual nodes in EMPIRE without power demand, and where it is possible to build either bottom-fixed or floating wind turbines. The modelled wind farm nodes are aggregations of all major offshore wind farm projects in the North Sea area~\cite{4COffshore}, and are grouped together based on proximity. It is assumed that all wind farms have a capacity density of 6 MW/km$^2$, following~\cite{Borrmann2018}. The wind farms are able to connect in the hybrid configuration, meaning that they can simultaneously connect to shore, to the energy hub and to other wind farms. In either case, the maximum capacity of each cable connecting to the wind farm is capped by the generation capacity of the wind farm. 

Furthermore, the maximum capacities of cable connections in the North Sea are limited. These limits are a maximum capacity of 1 GW between wind farms, 5 GW between a wind farm and its home country, 10 GW between a wind farm and the energy hub, and 20 GW between the energy hub and the North Sea countries. These limits are applied to all wind farms unless the existing capacity is higher, in which case the existing capacity is set as the maximum. 

The placement of this hub will impact on the results. It is therefore important to carefully choose the location of the energy hub.
A preliminary investigation with three potential locations was performed to determine the best location of the energy hub (\ie \ the location with the lowest system cost).  Figure \ref{fig:windfarms} shows the three locations, where the EU-focused hub (\ie, the one in the lower right) resulted in the lowest system cost, and was chosen for further analysis in this work. Note that hydrogen is only produced on the energy hub or onshore in the North Sea countries. That is, it is assumed that it is not viable to install electrolyzers on the offshore wind farms.

\FloatBarrier

\section{Results and analysis}

This section provides the main results from this study. Section \ref{sec:generatorcapacity} shows the growth in generation capacity in the North Sea wind farms and for Europe as a whole; Section \ref{sec:electrolyzercapacities} analyzes the evolution of the electrolyzer capacity in the North Sea area; Section \ref{sec:transmissioncapacity} shows the build-up of electricity transmission capacity; Section \ref{sec:curtailment} explores the effect of hydrogen and the energy hub on curtailment of North Sea wind farms and; Section \ref{sec:prices} analyzes the effects of hydrogen on power prices.

\subsection{Power generation capacity in Europe and the North Sea area}

A hydrogen economy based on electrolysis will increase the electricity demand which will drive investments into low-carbon power generators. The North Sea energy hub can be used to transmit energy among the North Sea countries and to produce hydrogen offshore. Placing electrolyzers on the North Sea energy hub may drive investments into the North Sea offshore wind farms. This section shows how the hydrogen economy and North Sea energy hub influence investments in the North Sea wind farms and the European power system as a whole.

\label{sec:generatorcapacity}

\subsubsection{Capacity of wind farm sites}
\label{sec:windfarmcapacities}

\begin{figure}[ht!]
    \centering
    \begin{subfigure}[b]{0.49\textwidth}
         \centering
         \includegraphics[width=\textwidth]{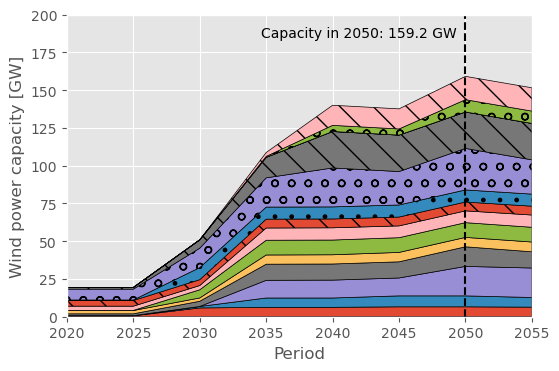}
         \caption{Without hydrogen \& with energy hub.}
         \label{fig:windfarmcapacity_hub_h20}
     \end{subfigure}
    \hfill
     \begin{subfigure}[b]{0.49\textwidth}
         \centering
         \includegraphics[width=\textwidth]{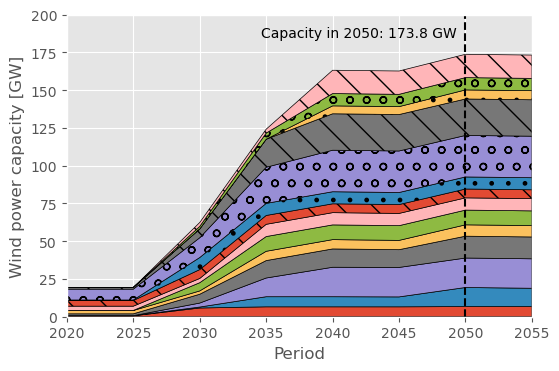}
         \caption{With hydrogen \& with energy hub.}
         \label{fig:windfarmcapacity_hub_h2100}
     \end{subfigure} \\
     \begin{subfigure}[b]{0.49\textwidth}
         \centering
         \includegraphics[width=\textwidth]{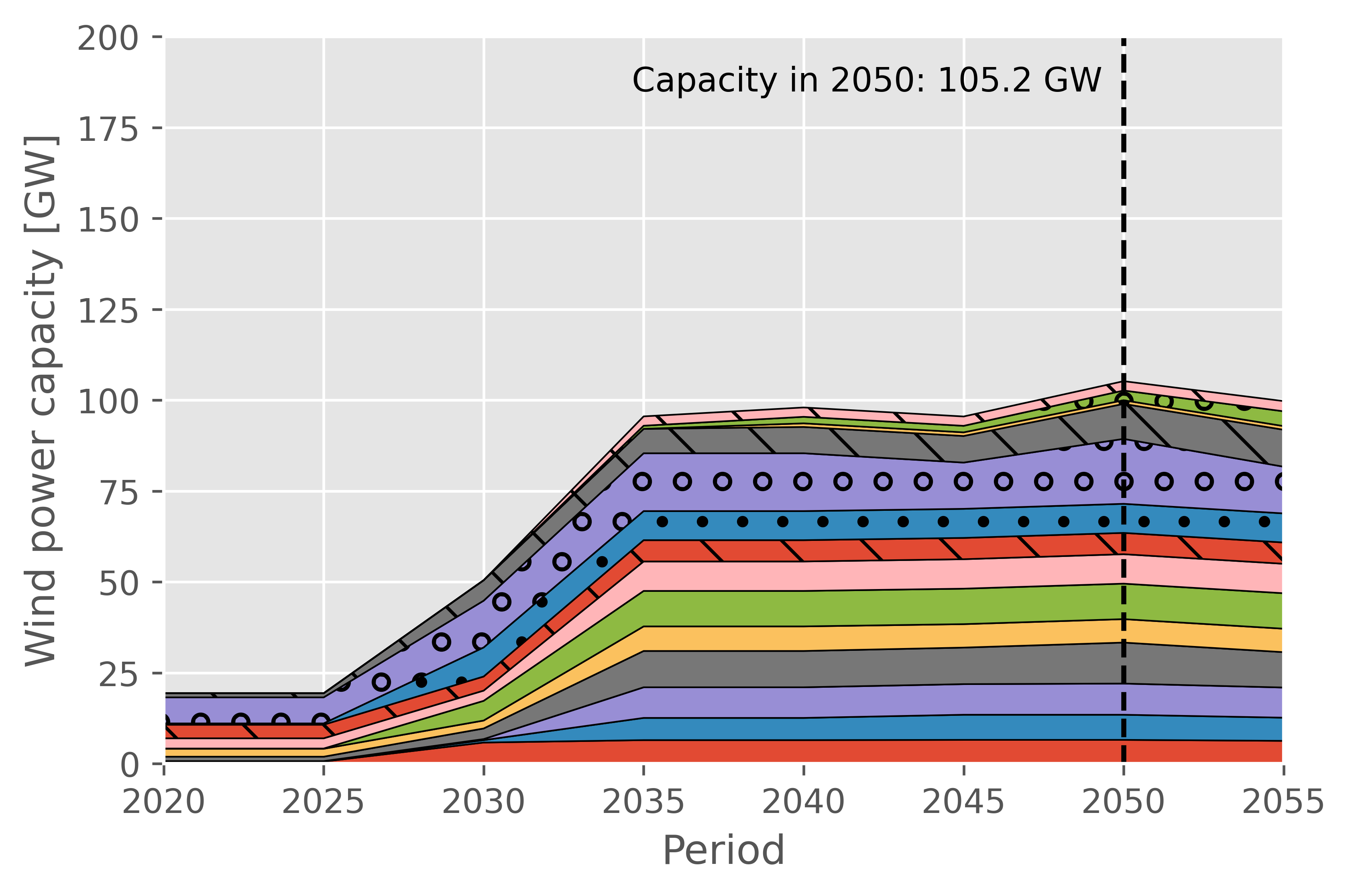}
         \caption{Without hydrogen \& without energy hub.}
         \label{fig:windfarmcapacity_nohub_h20}
     \end{subfigure} 
    \hfill
     \begin{subfigure}[b]{0.49\textwidth}
         \centering
         \includegraphics[width=\textwidth]{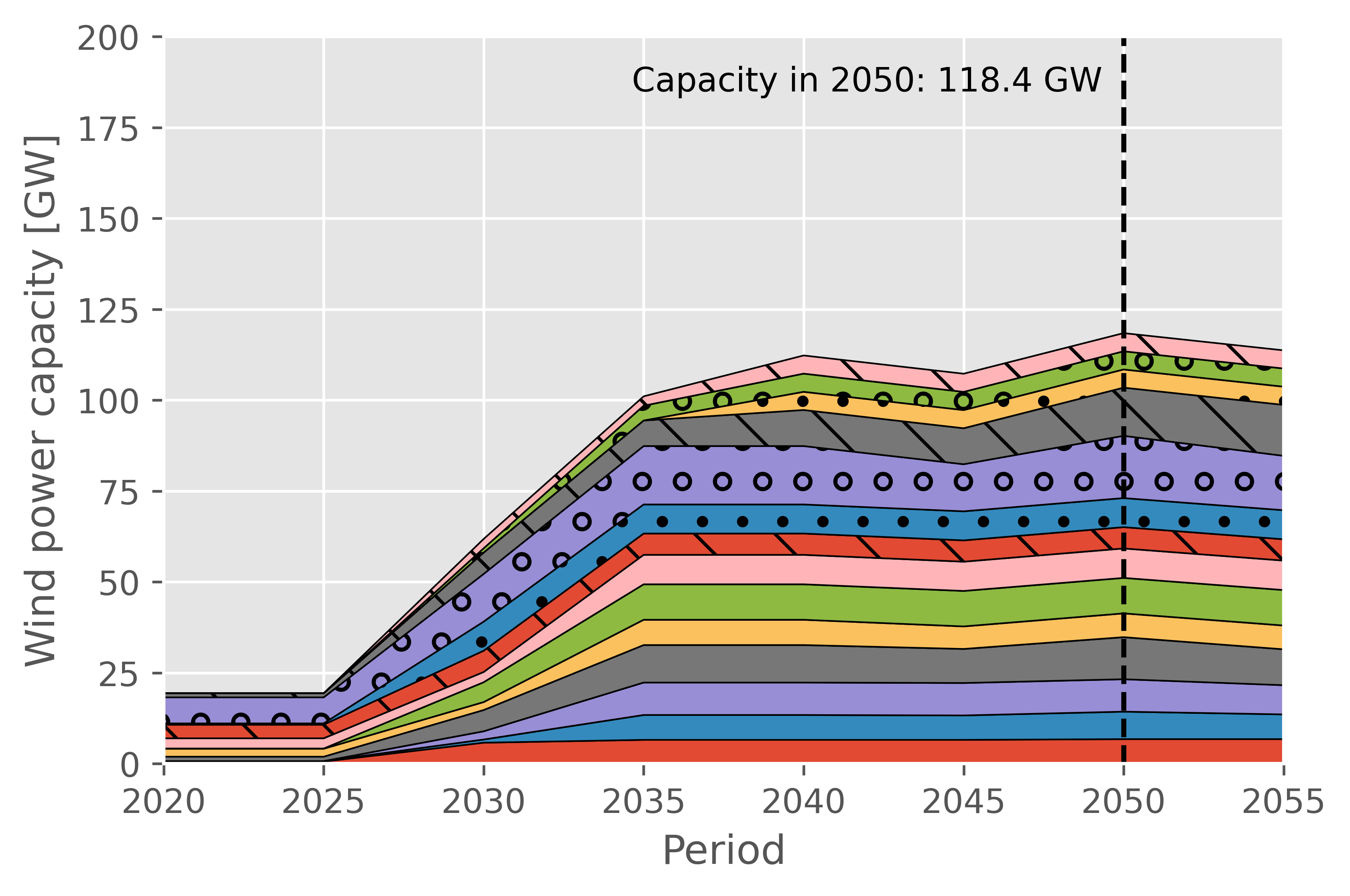}
         \caption{With hydrogen \& without energy hub.}
         \label{fig:windfarmcapacity_nohub_h2100}
     \end{subfigure} \\
     \begin{subfigure}[b]{0.7\textwidth}
         \centering
         \includegraphics[width=\textwidth]{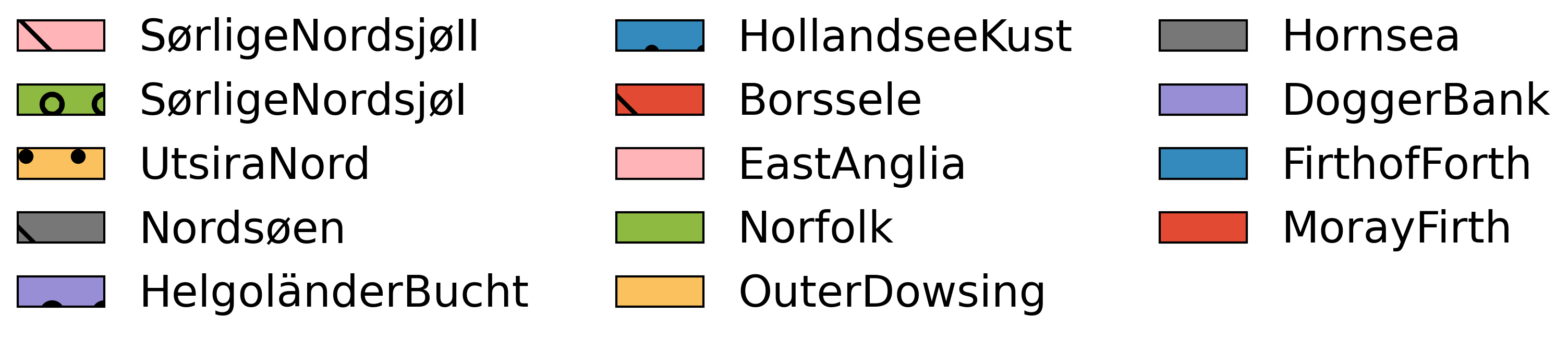}
     \end{subfigure} 
    \caption{Capacity of North Sea offshore wind farms.}
    \label{fig:offshorewindcapacities}
\end{figure}

Figure \ref{fig:offshorewindcapacities} shows how the capacities of the North Sea wind farms change with the introductions of the North Sea energy hub and the hydrogen economy. In the presence of the offshore energy hub, the total offshore wind power capacity in 2050 increases from 105.2 GW (Figure \ref{fig:windfarmcapacity_nohub_h20}) to 159.2 GW (Figure \ref{fig:windfarmcapacity_hub_h20}) -- an increase of approximately 51\%. When hydrogen production is considered the increase is from 118.4 GW (Figure \ref{fig:windfarmcapacity_nohub_h2100}) to 173.8 GW (Figure \ref{fig:windfarmcapacity_hub_h2100}), an increase of 47\%.

For some of the wind farms there is no difference between the four cases, such as \eg \  Moray Firth and Borssele. This is because in each case, the generation capacities of these wind farms are on their maximum limit. 
Other sites have their capacities significantly increased when either the energy hub or the hydrogen economy is included, such as the Nordsøen wind farm.  Other wind farm sites are sensitive to the cases, where \eg, the capacity of Utsira Nord is zero when only the energy hub is available, but is 5 GW when the hydrogen economy is included. Similarly, Dogger Bank's capacity increases only when the offshore energy hub is included, but not when only hydrogen is enabled.
All of these wind farms have an increased capacity when both the offshore energy hub and hydrogen are enabled, but it is also observed that some wind farms increase their capacity only when \textit{both} are added, \eg \  Firth of Forth, where the capacity increases to 12.6 GW when both the energy hub and hydrogen are enabled, up from approximately 7.0-7.5 GW in all other cases.

\FloatBarrier 

\subsubsection{Capacity mix in Europe}

\begin{figure}[ht!]
    \centering
    \begin{subfigure}[b]{0.49\textwidth}
         \centering
         \includegraphics[width=\textwidth]{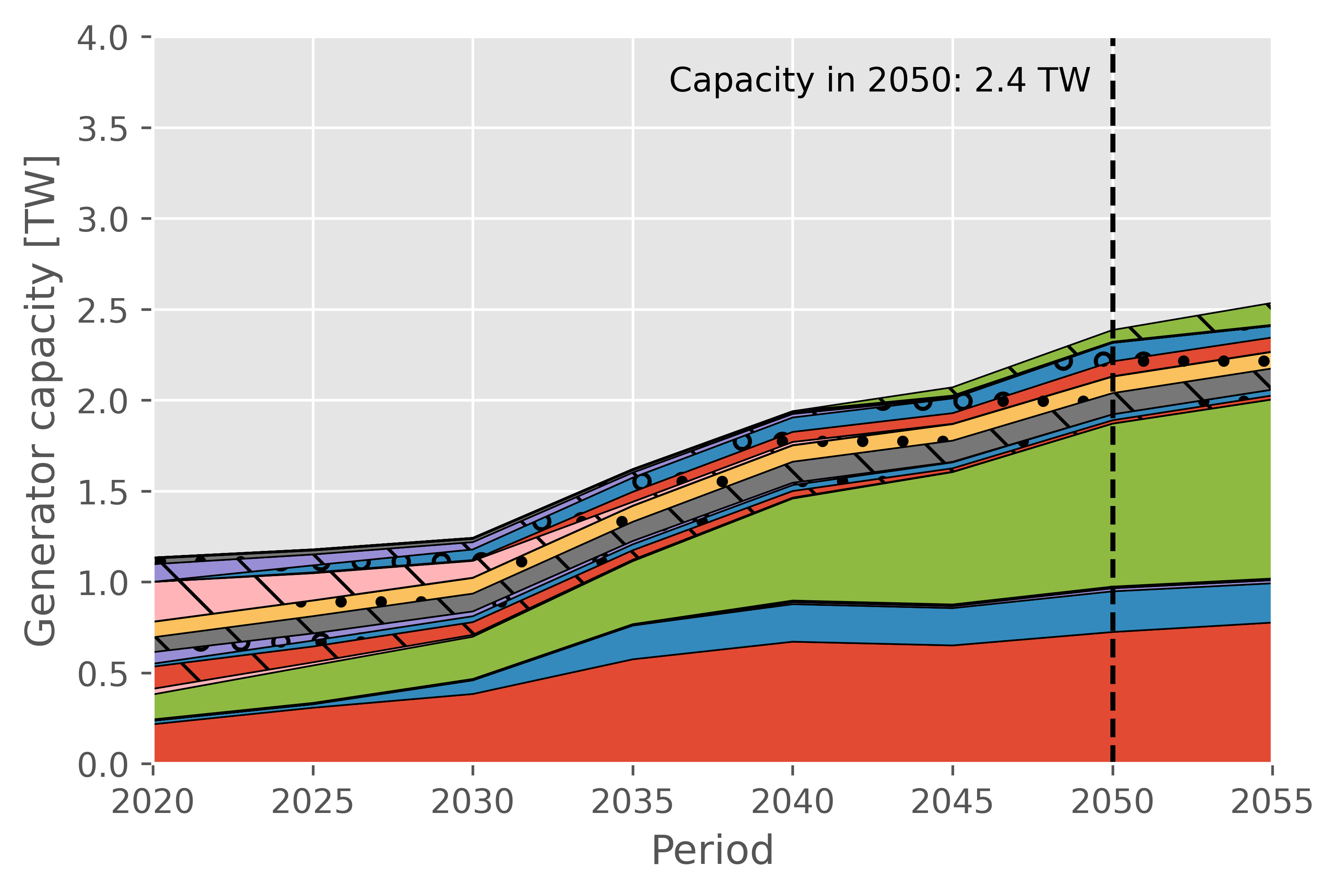}
         \caption{Without hydrogen \& with energy hub.}
         \label{fig:EUcapacity_hub_h20}
     \end{subfigure}
    \hfill
     \begin{subfigure}[b]{0.49\textwidth}
         \centering
         \includegraphics[width=\textwidth]{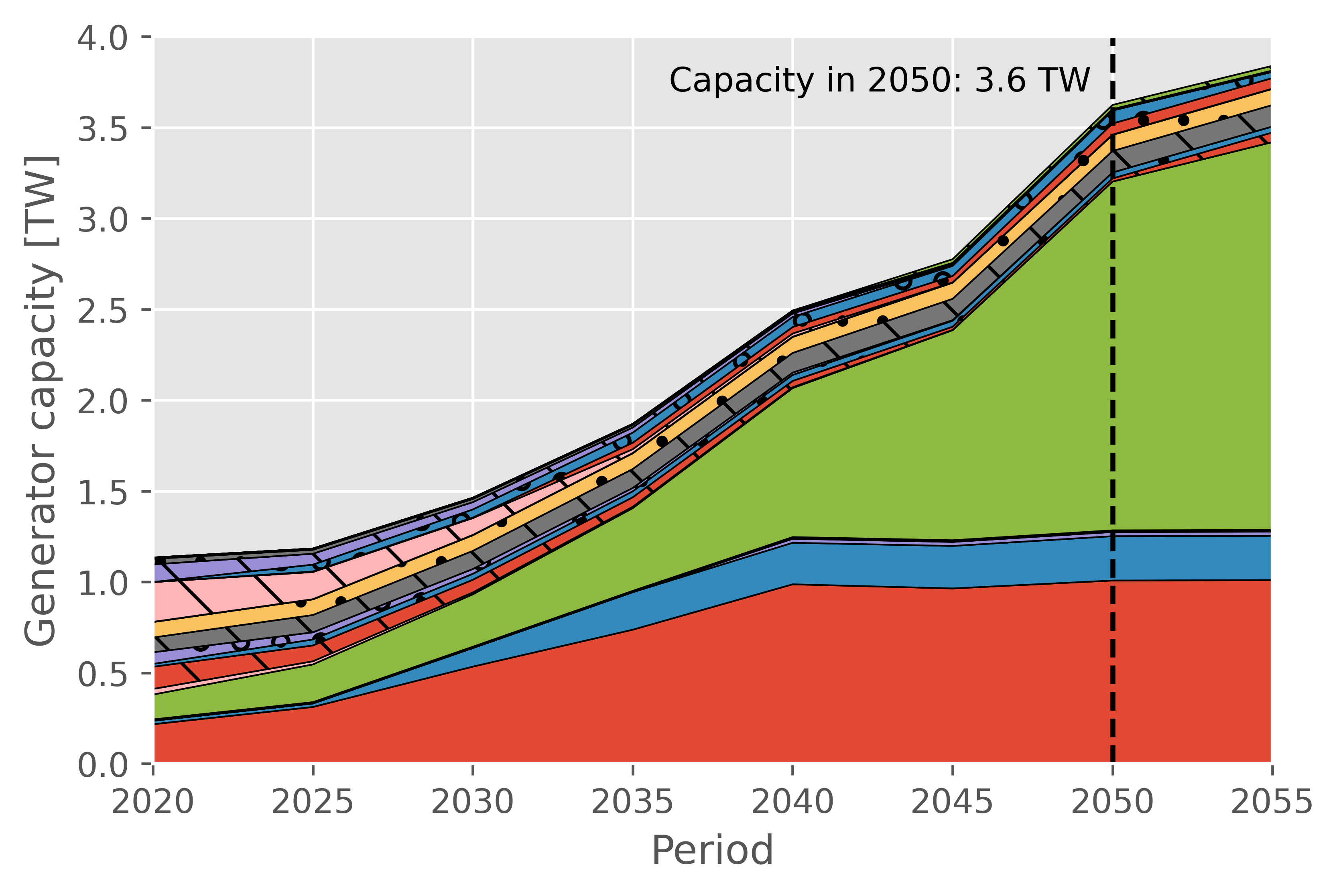}
         \caption{With hydrogen \& with energy hub.}
         \label{fig:EUcapacity_hub_h2100}
     \end{subfigure} \\
     \begin{subfigure}[b]{0.49\textwidth}
         \centering
         \includegraphics[width=\textwidth]{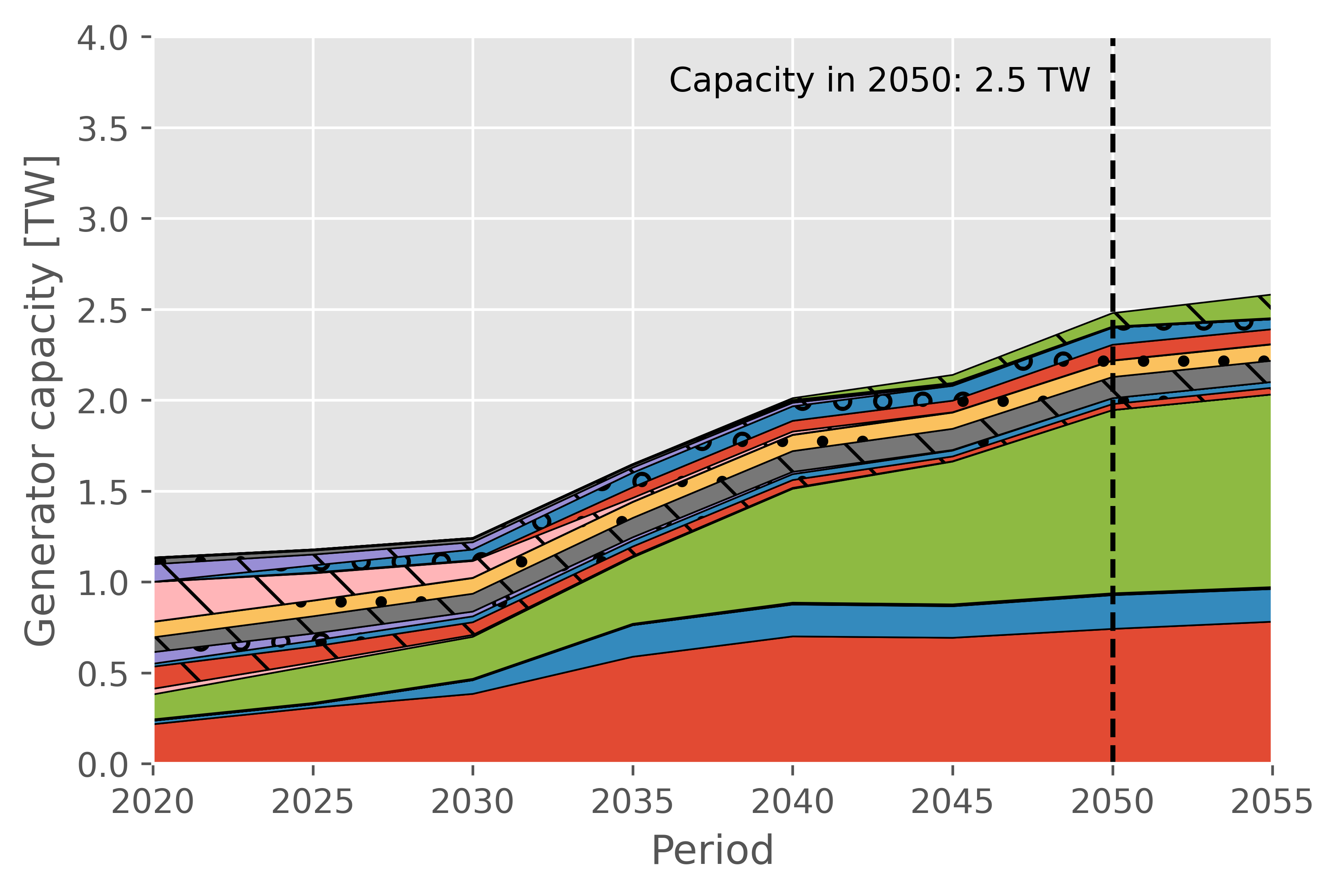}
         \caption{Without hydrogen \& without energy hub.}
         \label{fig:EUcapacity_nohub_h20}
     \end{subfigure} 
    \hfill
     \begin{subfigure}[b]{0.49\textwidth}
         \centering
         \includegraphics[width=\textwidth]{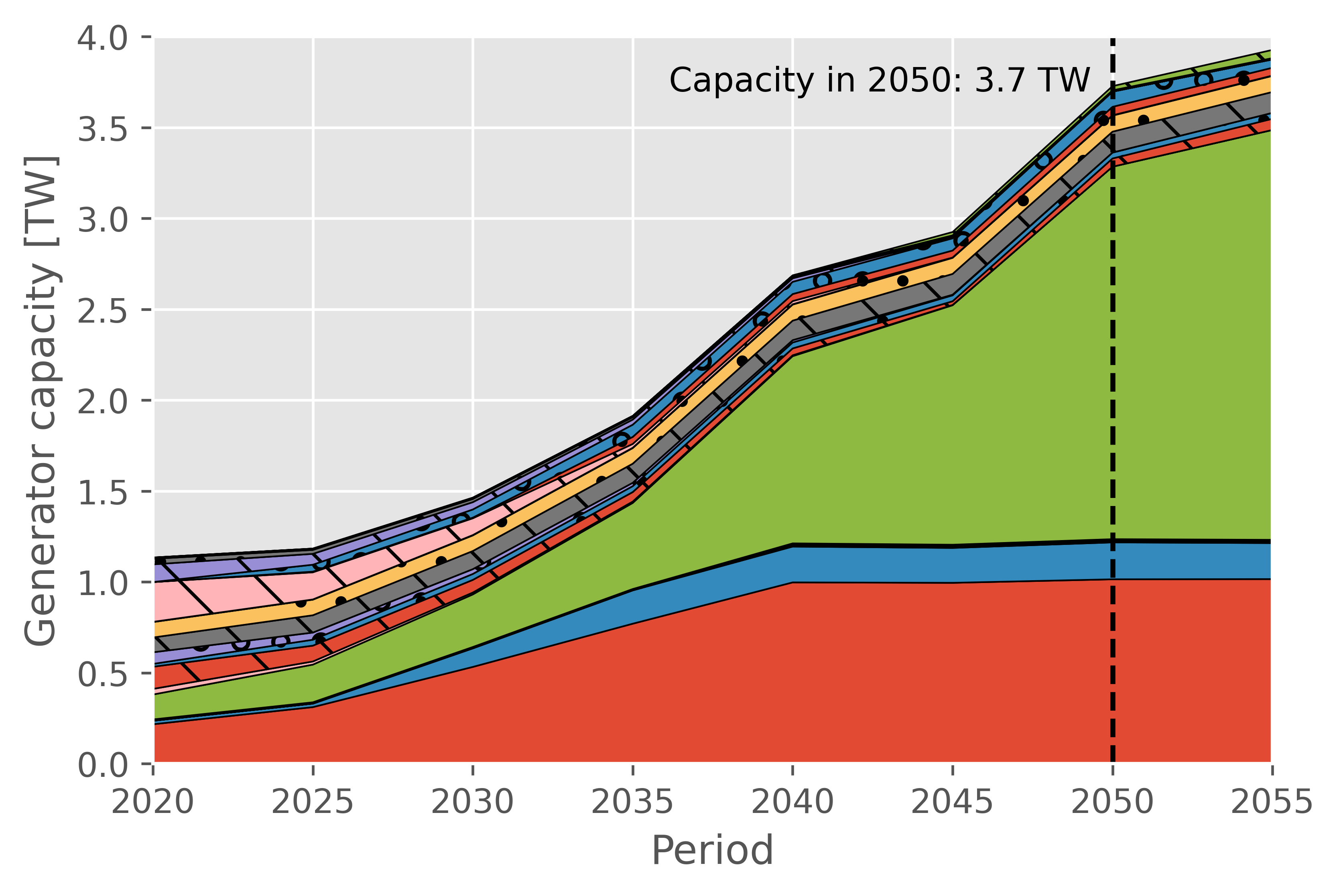}
         \caption{With hydrogen \& without energy hub.}
         \label{fig:EUcapacity_nohub_h2100}
     \end{subfigure} \\
     \begin{subfigure}[b]{0.8\textwidth}
         \centering
         \includegraphics[width=\textwidth]{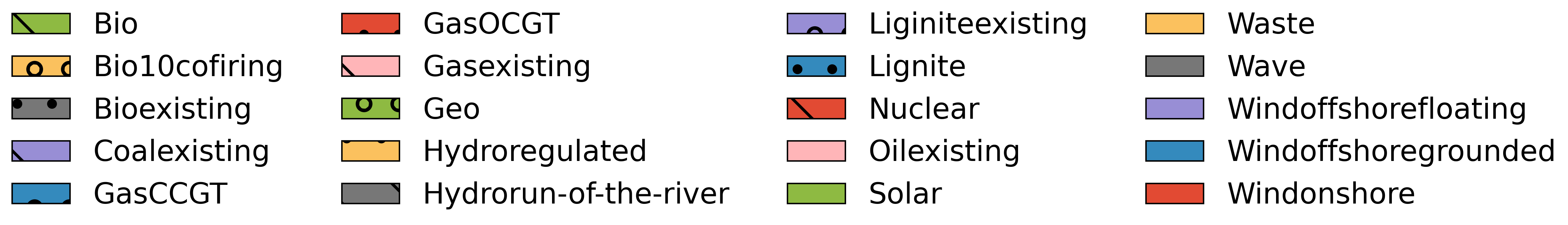}
     \end{subfigure} 
    \caption{Generation capacity of all generators in the European power market.}
    \label{fig:EU_generation_capacity}
\end{figure}

The overall power generation capacity mix in Europe is also affected by the inclusion of the offshore energy hub and hydrogen, see Figure \ref{fig:EU_generation_capacity}.

\FloatBarrier 

The inclusion of hydrogen leads to a considerable increase in total European generation capacity (Figures \ref{fig:EUcapacity_hub_h2100} and \ref{fig:EUcapacity_nohub_h2100}). Hydrogen is produced in Europe and exclusively through electrolysis, and this drives large investments into new generation power generation, increasing the total generation capacity in 2050 \ca \ 50\% with and without the energy hub.
The largest increases in generation capacity are for solar and wind power in order to achieve the decarbonization targets of the European Commission~\cite{EuropeanCommission2018}. This means that the hydrogen from 2040 forwards can be considered almost entirely green.

The inclusion of the offshore energy hub reduces the total European generation capacity in 2050 by 100 GW both with and without hydrogen. This creates changes in the distribution of generation technologies. Comparing Figures \ref{fig:EUcapacity_hub_h20} and \ref{fig:EUcapacity_nohub_h20}, the availability of the energy hub reduces the investments into solar by approximately 10\% (\ca \ 0.1 TW) while increasing the total capacity of offshore wind by 24\% (\ca \ 0.05 TW). In the absence of the offshore energy hub, there are no investments into floating offshore wind, but the capacity of floating offshore wind in the North Sea increases to approximately 0.02 TW once the energy hub is enabled. A similar trend occurs for the cases when hydrogen is enabled, showing how the transmission flexibility offered by the energy hub is valuable in a future power system that relies heavily on renewable energy sources. 



In short, the European energy transition with hydrogen requires large investments in energy generation. This goes against a commonly held belief that European hydrogen production will be driven by excess and otherwise curtailed renewable electricity production. Instead, the projected hydrogen demand will necessitate considerable investments in power generation, and if the power sector is to be decarbonized, most of these investments will be based on renewable energy sources, \ie, solar and wind.

\subsection{Electrolyzer capacities in the North Sea region}
\label{sec:electrolyzercapacities}

\begin{figure}[ht!]
    \centering
    \begin{subfigure}[b]{0.49\textwidth}
        \centering
        \includegraphics[width=\textwidth]{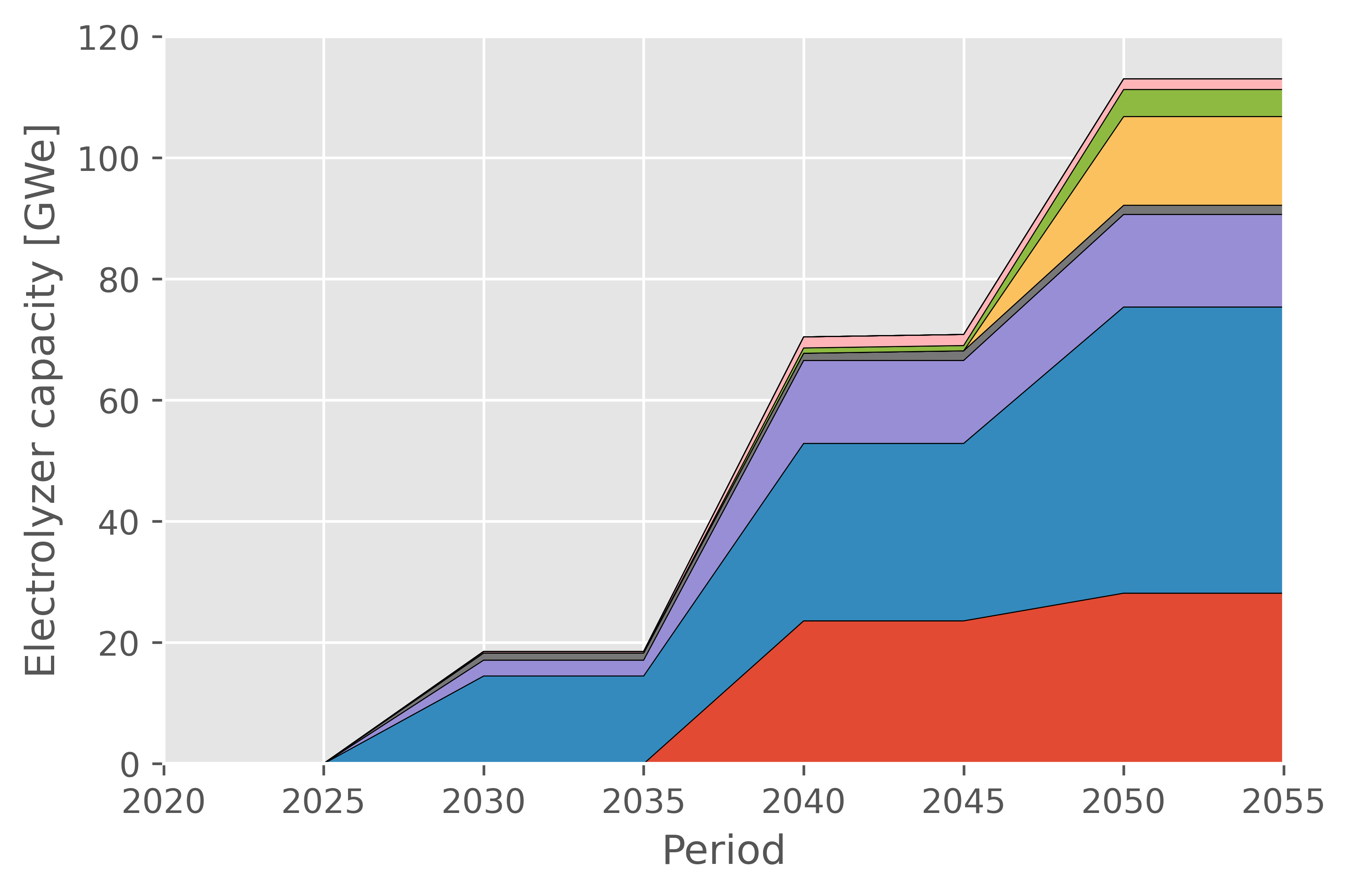}
        \caption{With energy hub.}
        \label{fig:electrolyzercapacity_hub}
    \end{subfigure}
    \hfill
    \begin{subfigure}[b]{0.49\textwidth}
        \centering
        \includegraphics[width=\textwidth]{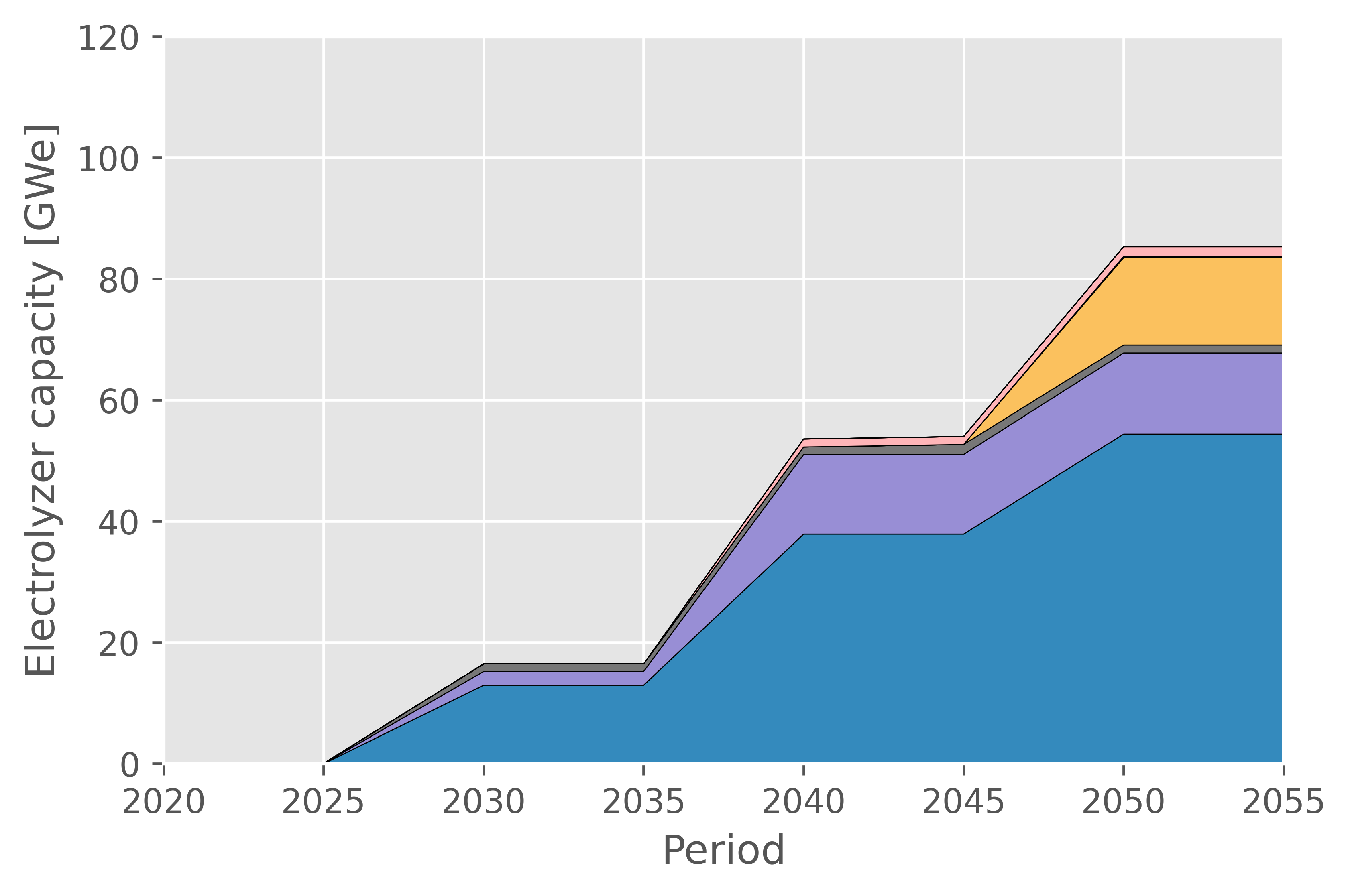}
        \caption{Without energy hub.}
        \label{fig:electrolyzercapacity_nohub}
    \end{subfigure} \\
    \begin{subfigure}[b]{0.7\textwidth}
        \centering
        \includegraphics[width=\textwidth]{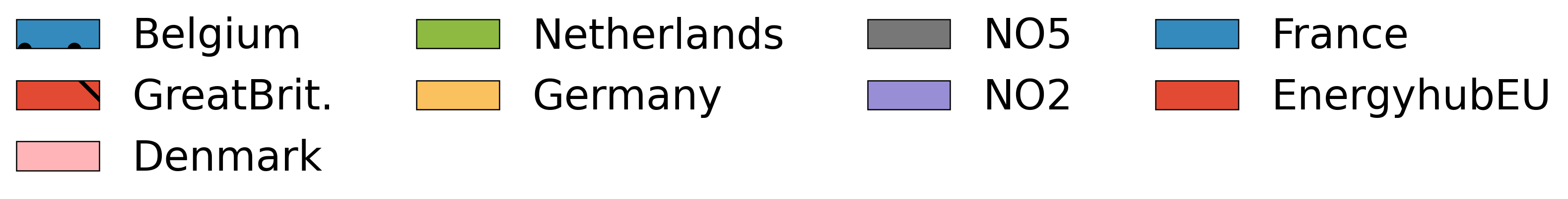}
    \end{subfigure}
    \caption{Capacity of electrolyzers in North Sea region. The unit is the electric capacity of the electrolyzers.}
    \label{fig:electrolyzercapacities}
\end{figure}

It is expected that the North Sea region will have a significant share of the total European electrolyzer capacity, given the favorable wind conditions in the North Sea,  Note that the hydrogen capacity assumes that all of the European hydrogen demand is satisfied by European production from electrolyzers. This means that the capacities reported here might be overestimated, as other hydrogen production methods (\eg, natural gas reforming) and imports are not considered. The European Commission has set a strategy for the build-up of green hydrogen production in the European Union (EU), and they have set a goal for 2030 of 40 GW of capacity in the EU and 40 GW in neighboring countries that will supply the EU with green hydrogen~\cite{europeancommission2020}. Thus, a total electrolyzer capacity of \ca \ 80 GW is expected in 2030, and for both cases presented in Figure \ref{fig:electrolyzercapacities}, the \textit{total European} capacity is approximately 72 GW, which is in line with the scale set by the European Commission.

\FloatBarrier 

In Figure \ref{fig:electrolyzercapacity_hub} it can be observed how the offshore energy hub has a significant share of the total electrolyzer capacity in the North Sea region. The energy hub is favored partly due to its proximity to the North Sea wind farms, and due to the interconnections through the energy hub. This means that the energy hub has access to copious renewable energy, enabling the hub to produce large amounts of green hydrogen. 

An important actor in the North Sea hydrogen production market is France which provides hydrogen to its large domestic demand while satisfying the demand of its neighbors in \eg, Germany and the UK. Additionally, France has a large deployment of low-carbon power generation technologies, including nuclear, solar and wind (both onshore and offshore), which allows France to produce large quantities of low-carbon hydrogen. This large domestic production is further supplemented by the large availability of renewable power in Spain and Germany.

Next to the offshore energy hub and France, in Figure \ref{fig:electrolyzercapacities} it can be observed how southern Norway, represented by the price zone NO2, and Germany are the remaining major producers of hydrogen in the North Sea region. Germany is a favorable location for electrolyzer capacity, as it can more easily trade with southern and eastern Europe. NO2 is primarily powered by domestic offshore wind (Utsira Nord, Sørlige Nordsjø I \& II) and by Norwegian hydropower. The availability of dispatchable and renewable power enables Norway to become a significant producer of European hydrogen. 



With the energy hub, the North Sea region accounts for approximately 25\% of the total European electrolyzer capacity in 2050. The country with the largest production is Spain, with approximately 85 GW of electrolyzer capacity. This hydrogen production is in large part powered by Spain's domestic production of solar power. Other producers follow the same pattern established in the North Sea region and in Spain: access to large quantities of renewable electricity facilitates the deployment of electrolyzers to satisfy European demand for low-carbon hydrogen.

In the case without the energy hub (Figure \ref{fig:electrolyzercapacity_nohub}), the  electrolyzer capacity in the North Sea region in 2050 decreases by approximately 25\%, accounting for approximately 19\% of the total European electrolyzer capacity. The decreased capacity roughly corresponds to the capacity that was located on the energy hub. Comparing Figures \ref{fig:electrolyzercapacity_nohub} and \ref{fig:electrolyzercapacity_hub}, it is seen that the Netherlands only has electrolyzer capacity when the energy hub is included. Once the energy hub is disabled, this capacity is instead redirected to France. The electrolyzer capacities in Germany and NO2 do not change significantly between these two cases.

\subsection{Power transmission capacities}
\label{sec:transmissioncapacity}
The wind farms serve two purposes: i) they generate CO2 emission-free electricity, and ii) they act as additional transmission points to connect the North Sea countries. Given the large wind farm capacities seen in Section \ref{sec:windfarmcapacities}, a large expansion of transmission capacity in the North Sea region is expected.

\begin{figure}[ht!]
    \centering
    \begin{subfigure}[b]{0.45\textwidth}
        \centering
        \includegraphics[width=\textwidth]{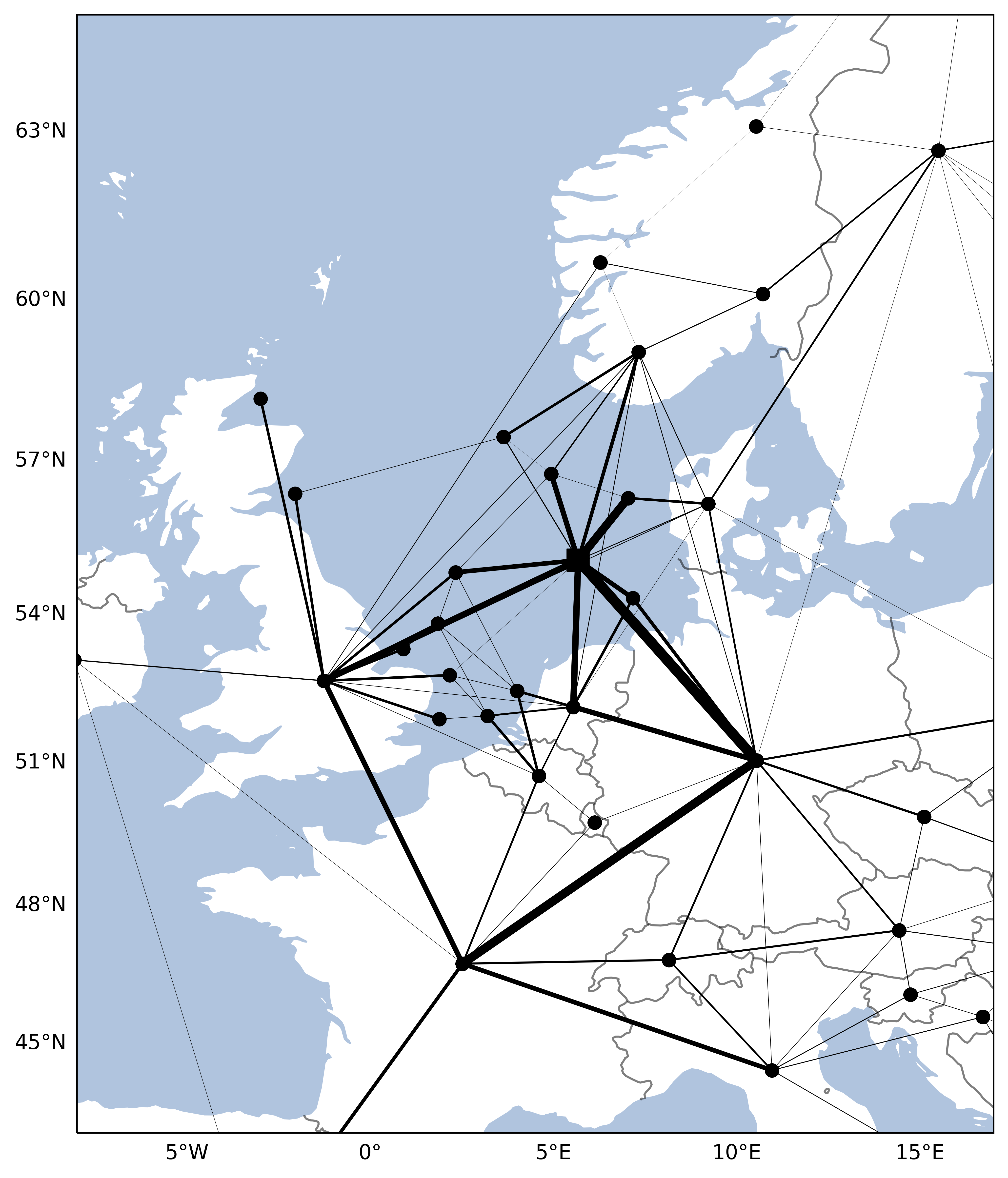}
        \caption{Without hydrogen \& with energy hub. Total capacity in North Sea: 248.7 GW.}
        \label{fig:tranmissioncapacity_hub_h20}
    \end{subfigure}
    \hfill
    \begin{subfigure}[b]{0.45\textwidth}
        \centering
        \includegraphics[width=\textwidth]{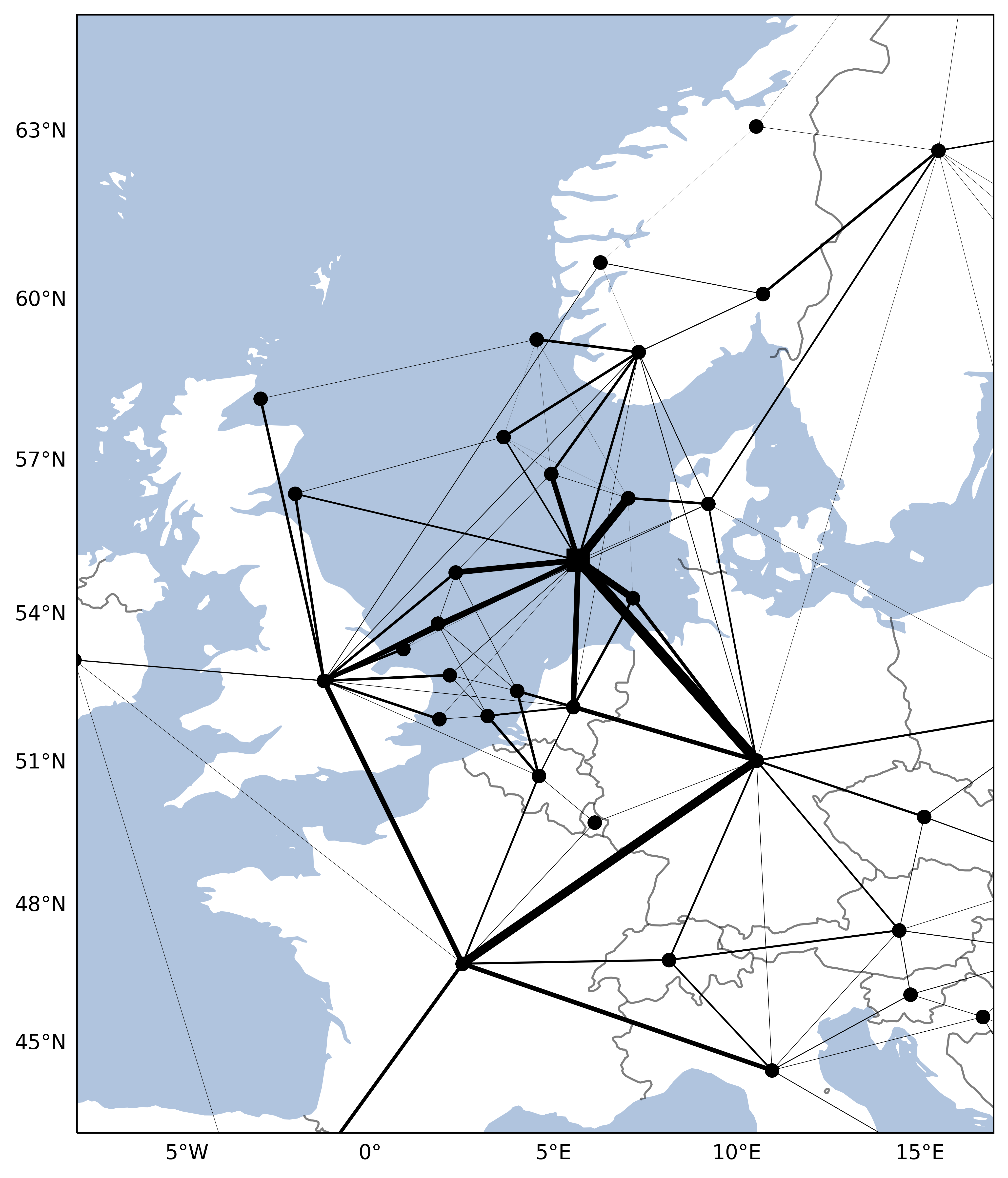}
        \caption{With hydrogen \& with energy hub. Total capacity in North Sea: 264.8 GW.}
        \label{fig:tranmissioncapacity_hub_h2100}
    \end{subfigure} \\
     \begin{subfigure}[b]{0.45\textwidth}
         \centering
         \includegraphics[width=\textwidth]{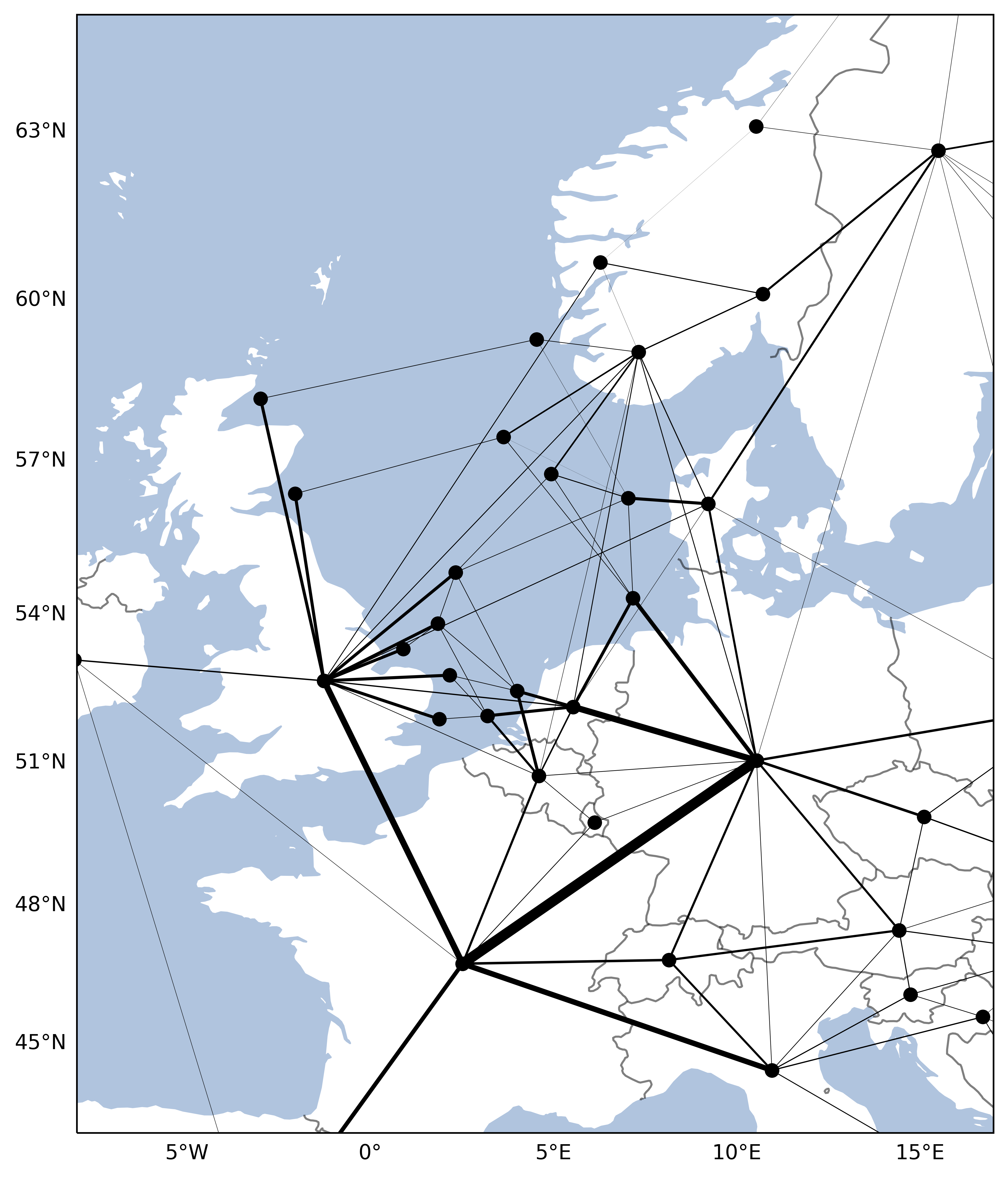}
         \caption{Without hydrogen \& without energy hub. Total capacity in North Sea: 154.0 GW.}
         \label{fig:tranmissioncapacity_nohub_h20}
     \end{subfigure} 
    \hfill
     \begin{subfigure}[b]{0.45\textwidth}
         \centering
         \includegraphics[width=\textwidth]{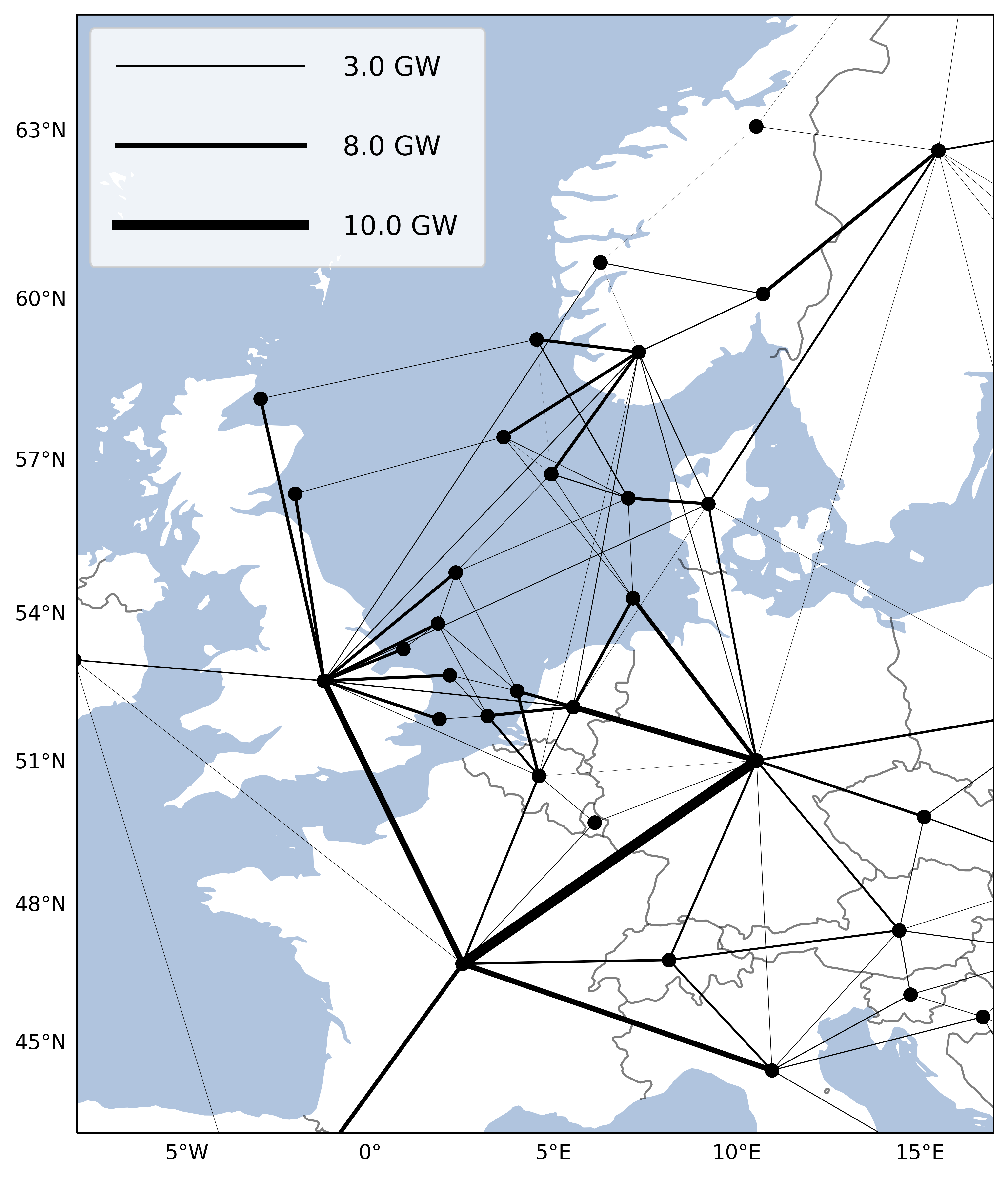}
         \caption{With hydrogen \& without energy hub. Total capacity in North Sea: 165.0 GW.}
         \label{fig:tranmissioncapacity_nohub_h2100}
     \end{subfigure}
    \caption{Transmission capacity in the North Sea region in 2050.}
    \label{fig:transmissioncapacities}
\end{figure}
\FloatBarrier 

In 2050, Figure \ref{fig:transmissioncapacities} depicts how the total transmission capacity in the North Sea increases significantly as hydrogen and the offshore energy hub are introduced. The total transmission capacity is considerably larger than today's grid in the North Sea, and a significant share is from connections between wind farms and from wind farms to shore. The total capacity numbers reported in Figure \ref{fig:transmissioncapacities} include capacities between North Sea countries as well (\eg, from Germany to the Netherlands). In Figure \ref{fig:tranmissioncapacity_nohub_h20}, the cable capacities to the wind farms account for approximately 82 GW of the total North Sea grid capacity. Related literature has also reported considerable expansions of North Sea grid capacities. For example,  \cite{hartel2021} predicts that the total transmission capacity in the North Sea  will be around 110 GW.

In the hydrogen only case, there is an increase in transmission capacities around the coast of Southern Norway (node NO2). There, the connections from the mainland to Utsira Nord, Sørlige Nordsjø I and Sørlige Nordsjø II are all increased. The transmission capacities to these three wind farms account for approximately 9 GW of the 11 GW increase in total interconnection capacity in the North Sea region. 

Introducing the offshore energy hub also leads to a substantial increase in transmission capacity. The capacity is a result of several North Sea countries and offshore wind farms connecting to the energy hub as a transmission asset. For the wind farms, the strongest connections are to the wind farm nodes closest to the European continent, \ie, Nordsøen by Denmark, Helgoländer Bucht by Germany, and Dogger Bank by the UK and Sørlige Nordsjø II by Norway. There is a strong connection between the energy hub and Germany, the Netherlands, the UK, Norway and Denmark. This suggests that the flexibility to transfer wind power and other electricity among the European continental countries is particularly valuable, as a 61\% increase in transmission capacity is observed in combination with a substantial (but smaller) increase in generation capacity among the offshore wind farms in this region.

Enabling both the offshore energy hub and hydrogen induces the largest increase in transmission capacity. Particularly, these increases in cable capacity are between the energy hub and wind farms that are far from the hub, including \eg \ Firth of Forth and Hornsea. The connections between the energy hub and these wind farms account for approximately 7 GW of the increase from Figure \ref{fig:tranmissioncapacity_hub_h20} to Figure \ref{fig:tranmissioncapacity_hub_h2100}. Other notable increases include the connection between the energy hub and Dogger Bank (\ca \ 2.3 GW), and between the energy hub and Helgoländer Bucht (\ca \ 5 GW). 

Overall, results indicate that a decarbonized European power system coupled with strong hydrogen demand requires considerable investments in the North Sea grid. The construction of an offshore energy hub will lead to a substantial grid development, where the energy hub enables large transmission between the offshore wind farms and between the North Sea countries. 


\subsection{Curtailments in the North Sea wind farms}
\label{sec:curtailment}

Both the offshore energy hub and hydrogen production have the potential to reduce curtailment from the North Sea wind farms. Note that in this paper, curtailment is defined as the share of intermittent renewable power that is not used in the power sector or for hydrogen production. Since the modelled hydrogen market does not fully cover all use cases of hydrogen, it is likely that the actual curtailment of these systems overestimated. The energy hub increases the amount of wind power that can be transmitted to the North Sea countries, while hydrogen production allows for the valorization of excess wind power. 
Comparing Figures \ref{fig:windfarmcapacity_hub_h2100} and \ref{fig:electrolyzercapacity_hub}, it can be observed that the increase in wind farm power generation capacity is significantly smaller than the investment in electrolyzer capacity in the North Sea region, meaning that each unit of electrolyzer capacity built in the North Sea countries does not necessitate an equal amount of North Sea wind power generation capacity. This suggests that existing European power generation infrastructure can be used to power new electrolyzer capacity, and requires sufficient transmission capacity.
Another important reason is that the North Sea wind farms are better utilized, as seen in Table \ref{tab:curtailment}. The most significant reduction in curtailment occurs when hydrogen is enabled: the curtailment is reduced from 24.9\% to 17.1\% when the energy hub is not enabled. These large reported curtailment numbers are in line with other system with as large shares of renewable power, as seen in Figure \ref{fig:EU_generation_capacity}~\cite{Troncoso2011ElectrolysersHydrogen}. In contrast, only a marginal reduction from 24.9\% to 24.2\% in curtailment when only the energy hub is enabled can be observed. When both the energy hub and hydrogen are enabled, then the curtailment is reduced to 9.6\%. This implies that excess wind power is used to produce green hydrogen, but only to a limited degree. Even with the hydrogen enabled, there is still significant degree of curtailment in the North Sea wind farms. This may in part be due to the incomplete modelling of the uses of the renewable energy. Other works that have also included \eg \ the heating market have reported a significant value to the adoption of renewable generators \cite{Mathiesen2015SmartSolutions,Connolly2016SmartUnion}, and a significant reduction in curtailment of renewable generation \cite{Thellufsen2020SmartContext}.

\begin{table}[ht!]
    \centering
    \begin{tabular}{lSS}
    \hline
        & \textbf{Without hydrogen} & \textbf{With hydrogen} \\
        \textbf{With energy hub} & 24.23\% & 9.62\%  \\
        \textbf{Without energy hub} & 24.85\% & 17.09\% \\ \hline
    \end{tabular}
    \caption{Average curtailment of all offshore wind farms in the North Sea in 2050.}
    \label{tab:curtailment}
\end{table}


\subsection{Influence of hydrogen on power prices}
\label{sec:prices}

As the projected hydrogen demand is high, the electrolyzers will consequently consume large amounts of electricity from the power grid. This will impact the power prices in Europe.
Table \ref{tab:powerprices_mean} shows the effect on power prices in southern Norway, Germany and France in each season. There, it is observed that summer prices increase, as electrolyzers use the cheap electricity to produce hydrogen. In contrast, the price in expensive seasons (\eg \ fall) are reduced  for Germany and France.
Altogether, this suggests that electrolyzers and the hydrogen economy have a stabilizing effect on the power prices.
The stabilizing effect of the hydrogen economy on the power market is made more apparent in the peak load periods in the European grid, represented by Peak 2 in Table \ref{tab:powerprices_mean}, where the price decreases significantly for all prize zones.


\begin{table}[ht!]
    \centering
    \makebox[\textwidth][c]{
    \begin{tabular}{lSSSSSS|S}
    \hline
        {\textbf{Price zone}} & {\textbf{Winter}} & {\textbf{Spring}} & {\textbf{Summer}} & {\textbf{Fall}} & {\textbf{Peak 1}} & {\textbf{Peak 2}} & {\textbf{Year average}} \\
        \hline
        & \multicolumn{7}{c}{\textbf{Without hydrogen}}\\
        \textbf{NO2} & 17.0 & 32.7 & 12.7 & 55.7 & 139.4 & 4 687.8 & 42.6\\
        \textbf{Germany} & 100.0 & 58.2 & 38.0 & 107.3 & 186.4 & 5 146.4 & 90.1 \\
        \textbf{France} & 86.9 & 43.0 & 26.7 & 89.4 & 182.4 & 5 355.5 & 76.3 \\
        & \multicolumn{7}{c}{\textbf{With hydrogen}}\\
        \textbf{NO2} & 65.4 & 39.7 & 62.6 & 77.5 & 63.7 & 658.3 & 63.0 \\
        \textbf{Germany} & 92.9 & 56.7 & 53.8 & 95.5 & 2 852.8 & 1 437.2 & 86.1 \\
        \textbf{France} & 84.0 & 46.3 & 51.7 & 83.2 & 2 851.8 & 1 485.1 & 77.8 \\
    \hline
    \end{tabular}
    }
    \caption{Mean of expected power prices (in \euro/MWh) for each season in southern Norway, Germany and France in 2050. The results are for the case without an energy hub.}
    \label{tab:powerprices_mean}
\end{table}
\FloatBarrier


This stabilizing effect is a result of the grid investments. The increased capacities of power generation and transmission assets (see Sections \ref{sec:generatorcapacity} and \ref{sec:transmissioncapacity})-- initially bought to power electrolyzers -- are used to satisfy load during high peak hours, rather than for hydrogen production. The grid investments for the hydrogen economy thus also increase the robustness of the European grid as well as reduced prices during times of peak demand. However, this price reduction is not ubiquitous. In Peak 2 (Table \ref{tab:powerprices_mean})  the power prices increase significantly. Here, the electrolyzer operations influence the price, as the electrolyzers produce hydrogen to meet the yearly demand. 

The impact on overall yearly price is not equal either. Southern Norway experiences an increase in yearly average expected power price, whereas Germany has its yearly average price slightly lowered. This is due to greater integration between the Norwegian and continental European power markets, originating from higher international transmission cable capacities, and resulting in more equal power prices. Norway's large increase is also due to a high utilization of its electrolyzer capacity, leading to considerably increased power demand.

Ultimately, the hydrogen economy has a significant impact on the power market. It is therefore clear that a widespread production of hydrogen in Europe will not primarily use otherwise curtailed renewable power. This impact might have a positive side as the increase in generation capacity will support the power grid during peak load hours.

\section{Conclusion}

This work has investigated how the construction of an energy hub in the North Sea and the future European hydrogen economy affect the development of the North Sea grid. The curtailment of the offshore wind power from the North Sea is at most reduced from 24.9\% to 9.6\% by the inclusion of the energy hub and hydrogen production. In the hydrogen sector, it is shown how electrolyzer capacities are deployed in the North Sea region and their influences on power prices. This leads to the following insights:

\begin{itemize}
    \item \textbf{The inclusion of the energy hub facilitates the increase in offshore wind capacity in the North Sea.} By facilitating more energy transmission between the North Sea wind farms and nearby countries, the energy hub increases the value of the North Sea wind resources. The transmission capacity expansion allows for a significant reduction in total European power generation capacity. 
    \item \textbf{A large European deployment of green hydrogen production will lead to a significant increase in European power generation.} Additionally, the increased power demand from electrolyzers will drive investments into North Sea wind power.
    \item \textbf{France, Norway and Germany are major producers of hydrogen in the North Sea region.} Of these, Norway consistently has the highest capacity factor, owing to its reduced interconnection with the European continent and the availability of hydropower and domestic offshore wind.
    \item \textbf{The production of green hydrogen can significantly lower the curtailment off North Sea offshore wind.} This holds particularly if there is sufficiently large interconnection capacity (through \eg \ the energy hub). Electrolyzers utilize some of the excess renewable energy, but green hydrogen production is not sufficient to avoid curtailment.
    \item \textbf{The deployment of the hydrogen economy has a significant impact on the power prices in Norway, France and Germany.} In Norway, this leads to an substantial increase in yearly average expected power prices, where the largest increases are seen in the seasons where the price is typically low (\eg \ summer). France and Germany also experience considerable price increases during otherwise low-price seasons, but there is no increase in yearly average power price.
    This is a result of how the grid investments made primarily for hydrogen generation (\ie, increased transmission and generation capacity) have positive effects also in the power market itself. 
    This is particularly evident during periods of peak demand.
    \item \textbf{The deployment of large-scale hydrogen production cannot primarily rely on otherwise curtailed renewable generation.} The green hydrogen economy will require significant investments in (renewable) power generation and transmission. Despite this, a substantial reduction of curtailment is observed, and this effect of hydrogen production should not be entirely discounted.
\end{itemize}

Based on above insights, further research should consider exploring: 

\begin{itemize}
    \item \textbf{Better handling of endogenous hydrogen demand.} Hydrogen can be used as a low-carbon industrial feedstock, or as an energy carrier in the power, transport and heat energy sectors. To better understand the value of hydrogen, this demand has to be endogenized in the capacity expansion model, rather than be set exogenously. This would allow for a deeper investigation into how electricity, hydrogen and bio-energy interact to satisfy the energy needs in industry, transport and the heat sector, where there is tremendous potential for synergies~\cite{Sorkns2020SmartMarkets}.
    \item \textbf{National versus European strategies.} National energy policy and strategies should be looked at more closely, and maybe imposed in the modelling analysis. The model optimizes an overall European Energy Transition perspective on the deployment of technologies. This at times might be at odds with some planned infrastructure projects or expected developments at the national or regional level.
    \item \textbf{Include blue hydrogen.} Blue hydrogen can be one of the initial ways of producing large volumes of cheap low-carbon hydrogen, and this can potentially be used to construct necessary infrastructure that can later be used by the green hydrogen economy. 
\end{itemize}

\section*{Acknowledgments}
 Thanks to the research grant (308811) in the project:  Planning Clean Energy Export from Norway to Europe. The authors acknowledge the financial support from the Research Council of Norway and the user partners Agder Energi, Air Liquide, Equinor Energy, Gassco and TotalEnergies E\&P Norge.
 
\section*{Appendix - Nomenclature}

\begin{longtable}{p{0.3\linewidth} p{0.65\linewidth}} 
    \textit{Sets} & \\
        $\mathscr{I}$ & periods  \\
        $\mathscr{S}$ & seasons \\
        $\mathscr{H}$ & operational hours \\
        $\Omega$ & set of scenarios\\
        $\mathscr{A}^{hyd}_n$ & set of nodes connected to node $n$ by hydrogen pipelines \\
        $\mathscr{N}^{hyd}$ & hydrogen production nodes \\
        $\mathscr{N}^{hub}$ & set of nodes that are offshore hubs \\
        $\mathscr{L}$ & all bidirectional power transmission arcs \\
        $\mathscr{L}_n$ & all bidirectional power transmission arcs connected to node $n$ \\
        $\mathscr{L}^{hyd}$ & bidirectional hydrogen pipes \\[2mm]
    
    \textit{Parameters} & \\
        $C^{el}_{i}$ & electrolyzer investment cost in period $i$, $\frac{\textup{\euro}}{MW (el)}$ \\
        $C^{pipe}_{l,i}$ & hydrogen pipeline investment cost for link $l$ in period $i$, $\frac{\textup{\euro}}{kg}$ \\
        $\Bar{C}^{storage}$ & levelized cost of hydrogen storage, $\frac{\textup{\euro}}{kg}$\\
        $L^{period}$ & length of periods, years\\
        $\varphi$ & operational discount rate, - \\
        $r$ & discount rate, - \\
        $pi_\omega$&  probability of scenario $\omega$, -  \\
        $\alpha_s$ & Seasonal scaling factor\\
        $LHV^{H_2}$ & Lower heating value of hydrogen, $\frac{MWh}{kg}$ \\
        $\eta^{p2h}_i$ & Power consumption to produce one kg of H$_2$ in period  $i$, $\frac{MWh (el.)}{kg}$\\
        $\bar{x}^{a}_{n,i}$ & remaining available capacity of initial capacity for asset $a$ in node $n$ in period $i$\\
        $i^{life}_a$& lifetime of asset $a$, years\\[2mm]
    
    \textit{Investment variables} & \\
        $x^{el}_{n,i}$ & Investment in electrolyzer in node $n$ in period $i$, MW\\
        $x^{pipe}_{l,i}$ & Investment in hydrogen pipe on link $l$ in period $i$, kg\\
        $x^{conv}_{n,i}$ & Investment in electrical converter in offshore hub\\[2mm]
    \textit{Operational variables} & \\
        $y^{p2h}_{n,h,i,\omega}$ & power used for hydrogen production in node $n$ in operational hour $h$ in period $i$ in scenario $\omega$ \\
        $y^{hyd}_{n,h,i,\omega} \leq v^{el}_{n,i}$ & hydrogen produced in node $n$ in operational hour $h$ in period $i$ in scenario $\omega$, kg \\
        $y^{hyd,trans}_{n,n_2,h,i,\omega} \leq v^{hyd,trans}_{n,n2,i}$ & hydrogen transported for node $n$ to node $n_2$ in operational hour $h$ in period $i$ in scenario $\omega$ \\
        $y^{hyd,sold}_{n,h,i,\omega}$& hydrogen sold in node $n$ in operational hour $h$ in period $i$ in scenario $\omega$, kg \\
        $y^{hyd,h2p}_{n,h,i,\omega}$ & hydrogen used for power production in node $n$ in operational hour $h$ in period $i$ in scenario $\omega$ \\
\end{longtable}

\setlength{\bibsep}{0pt plus 0.3ex} 
\footnotesize{ 
\bibliographystyle{ieeetr}
\bibliography{bibliography}} 

\begin{thebibliography}{10}

\bibitem{EuropeanCommission2018}
{European Commission}, ``{A Clean Planet for all: A European long-term
  strategic vision for a prosperous, modern, competitive and climate neutral
  economy},'' November 2018.

\bibitem{europeancommissionWind2020}
{European Commission}, ``{An EU Strategy to harness the potential of offshore
  renewable energy for a climate neutral future},'' November 2020.

\bibitem{europeancommission2020}
{European Commission}, ``{A hydrogen strategy for a climate-neutral Europe},''
  July 2020.

\bibitem{Yoshida2022}
A.~Yoshida, H.~Nakazawa, N.~Kenmotsu, and Y.~Amano, ``{Economic analysis of a
  proton exchange membrane electrolyser cell for hydrogen supply scenarios in
  Japan},'' {\em Energy}, vol.~251, p.~123943, 7 2022.

\bibitem{Karayel2022}
G.~K. Karayel, N.~Javani, and I.~Dincer, ``{Effective use of geothermal energy
  for hydrogen production: A comprehensive application},'' {\em Energy},
  vol.~249, p.~123597, 6 2022.

\bibitem{meier2014}
K.~Meier, ``{Hydrogen production with sea water electrolysis using Norwegian
  offshore wind energy potentials},'' {\em International Journal of Energy and
  Environmental Engineering}, vol.~5, p.~104, 2014.

\bibitem{babarit2018}
A.~Babarit, J.~C. Gilloteaux, G.~Clodic, M.~Duchet, A.~Simoneau, and M.~F.
  Platzer, ``{Techno-economic feasibility of fleets of far offshore
  hydrogen-producing wind energy converters},'' {\em International Journal of
  Hydrogen Energy}, vol.~43, pp.~7266--7289, 4 2018.

\bibitem{woznicki2020}
M.~Woznicki, G.~L. Solliec, and R.~Loisel, ``{Far off-shore wind energy-based
  hydrogen production: Technological assessment and market valuation
  designs},'' {\em Journal of Physics: Conference Series}, vol.~1669,
  p.~012004, 2020.

\bibitem{yan2021}
Y.~Yan, H.~Zhang, Q.~Liao, Y.~Liang, and J.~Yan, ``{Roadmap to hybrid offshore
  system with hydrogen and power co-generation},'' {\em Energy Conversion and
  Management}, vol.~247, p.~114690, 11 2021.

\bibitem{damore2021}
R.~d'Amore Domenech, T.~J. Leo, and B.~G. Pollet, ``{Bulk power transmission at
  sea: Life cycle cost comparison of electricity and hydrogen as energy
  vectors},'' {\em Applied Energy}, vol.~288, p.~116625, 4 2021.

\bibitem{singlitico2021}
A.~Singlitico, J.~Østergaard, and S.~Chatzivasileiadis, ``{Onshore, offshore
  or in-turbine electrolysis? Techno-economic overview of alternative
  integration designs for green hydrogen production into Offshore Wind Power
  Hubs},'' {\em Renewable and Sustainable Energy Transition}, vol.~1,
  p.~100005, 8 2021.

\bibitem{Calado2021b}
G.~Calado and R.~Castro, ``{Hydrogen Production from Offshore Wind Parks:
  Current Situation and Future Perspectives},'' {\em Applied Sciences},
  vol.~11, p.~5561, 6 2021.

\bibitem{andre2013}
J.~André, S.~Auray, J.~Brac, D.~D. Wolf, G.~Maisonnier, M.-M. Ould-Sidi, and
  A.~Simonnet, ``{Design and dimensioning of hydrogen transmission pipeline
  networks},'' {\em European Journal of Operational Research}, vol.~229,
  pp.~239--251, 8 2013.

\bibitem{baufume2013}
S.~Baufumé, F.~Grüger, T.~Grube, D.~Krieg, J.~Linssen, M.~Weber, J.-F. Hake,
  and D.~Stolten, ``{GIS-based scenario calculations for a nationwide German
  hydrogen pipeline infrastructure},'' {\em International Journal of Hydrogen
  Energy}, vol.~38, pp.~3813--3829, 4 2013.

\bibitem{reuss2019}
M.~Reu{\ss}, L.~Welder, J.~Thürauf, J.~Lin{\ss}en, T.~Grube, L.~Schewe,
  M.~Schmidt, D.~Stolten, and M.~Robinius, ``{Modeling hydrogen networks for
  future energy systems: A comparison of linear and nonlinear approaches},''
  {\em International Journal of Hydrogen Energy}, vol.~44, pp.~32136--32150, 12
  2019.

\bibitem{almansoori2006}
A.~Almansoori and N.~Shah, ``{Design and Operation of a Future Hydrogen Supply
  Chain},'' {\em Chemical Engineering Research and Design}, vol.~84,
  pp.~423--438, 6 2006.

\bibitem{kim2008b}
J.~Kim and I.~Moon, ``{Strategic design of hydrogen infrastructure considering
  cost and safety using multiobjective optimization},'' {\em International
  Journal of Hydrogen Energy}, vol.~33, pp.~5887--5896, 11 2008.

\bibitem{dayhim2014}
M.~Dayhim, M.~A. Jafari, and M.~Mazurek, ``{Planning sustainable hydrogen
  supply chain infrastructure with uncertain demand},'' {\em International
  Journal of Hydrogen Energy}, vol.~39, pp.~6789--6801, 4 2014.

\bibitem{kim2017}
M.~Kim and J.~Kim, ``{An integrated decision support model for design and
  operation of a wind-based hydrogen supply system},'' {\em International
  Journal of Hydrogen Energy}, vol.~42, pp.~3899--3915, 2 2017.

\bibitem{Almansoori2009}
A.~Almansoori and N.~Shah, ``{Design and operation of a future hydrogen supply
  chain: Multi-period model},'' {\em International Journal of Hydrogen Energy},
  vol.~34, pp.~7883--7897, 10 2009.

\bibitem{kim2016}
M.~Kim and J.~Kim, ``{Optimization model for the design and analysis of an
  integrated renewable hydrogen supply (IRHS) system: Application to Korea's
  hydrogen economy},'' {\em International Journal of Hydrogen Energy}, vol.~41,
  pp.~16613--16626, 10 2016.

\bibitem{nunes2015}
P.~Nunes, F.~Oliveira, S.~Hamacher, and A.~Almansoori, ``{Design of a hydrogen
  supply chain with uncertainty},'' {\em International Journal of Hydrogen
  Energy}, vol.~40, pp.~16408--16418, 12 2015.

\bibitem{kim2008a}
J.~Kim, Y.~Lee, and I.~Moon, ``{Optimization of a hydrogen supply chain under
  demand uncertainty},'' {\em International Journal of Hydrogen Energy},
  vol.~33, pp.~4715--4729, 9 2008.

\bibitem{tlili2019}
O.~Tlili, C.~Mansilla, M.~Robinius, K.~Syranidis, M.~Reu{\ss}, J.~Lin{\ss}en,
  J.~André, Y.~Perez, and D.~Stolten, ``{Role of electricity interconnections
  and impact of the geographical scale on the French potential of producing
  hydrogen via electricity surplus by 2035},'' {\em Energy}, vol.~172,
  pp.~977--990, 4 2019.

\bibitem{pan2020}
G.~Pan, W.~Gu, H.~Qiu, Y.~Lu, S.~Zhou, and Z.~Wu, ``{Bi-level mixed-integer
  planning for electricity-hydrogen integrated energy system considering
  levelized cost of hydrogen},'' {\em Applied Energy}, vol.~270, p.~115176, 7
  2020.

\bibitem{xiong2021}
B.~Xiong, J.~Predel, P.~C. del Granado, and R.~Egging-Bratseth, ``{Spatial
  flexibility in redispatch: Supporting low carbon energy systems with
  Power-to-Gas},'' {\em Applied Energy}, vol.~283, p.~116201, 2 2021.

\bibitem{agnolucci2013}
P.~Agnolucci and W.~McDowall, ``{Designing future hydrogen infrastructure:
  Insights from analysis at different spatial scales},'' {\em International
  Journal of Hydrogen Energy}, vol.~38, pp.~5181--5191, 5 2013.

\bibitem{Li2019}
L.~Li, H.~Manier, and M.-A. Manier, ``{Hydrogen supply chain network design: An
  optimization-oriented review},'' {\em Renewable and Sustainable Energy
  Reviews}, vol.~103, pp.~342--360, 4 2019.

\bibitem{FODSTAD2022112246}
M.~Fodstad, P.~{Crespo del Granado}, L.~Hellemo, B.~R. Knudsen, P.~Pisciella,
  A.~Silvast, C.~Bordin, S.~Schmidt, and J.~Straus, ``{Next frontiers in energy
  system modelling: A review on challenges and the state of the art},'' {\em
  Renewable and Sustainable Energy Reviews}, vol.~160, p.~112246, 2022.

\bibitem{almansoori2012}
A.~Almansoori and N.~Shah, ``{Design and operation of a stochastic hydrogen
  supply chain network under demand uncertainty},'' {\em International Journal
  of Hydrogen Energy}, vol.~37, pp.~3965--3977, 3 2012.

\bibitem{weimann2021}
L.~Weimann, P.~Gabrielli, A.~Boldrini, G.~J. Kramer, and M.~Gazzani, ``{Optimal
  hydrogen production in a wind-dominated zero-emission energy system},'' {\em
  Advances in Applied Energy}, p.~100032, 5 2021.

\bibitem{greiner2007}
C.~J. Greiner, M.~Korpås, and A.~T. Holen, ``{A Norwegian case study on the
  production of hydrogen from wind power},'' {\em International Journal of
  Hydrogen Energy}, vol.~32, pp.~1500--1507, 7 2007.

\bibitem{Backe2022}
S.~Backe, C.~Skar, P.~C. del Granado, O.~Turgut, and A.~Tomasgard, ``{EMPIRE:
  An open-source model based on multi-horizon programming for energy transition
  analyses},'' {\em SoftwareX}, vol.~17, p.~100877, January 2022.

\bibitem{Skar2016b}
C.~Skar, G.~L. Doorman, G.~A. P{\'{e}}rez-Vald{\'{e}}s, and A.~Tomasgard, ``{A
  multi-horizon stochastic programming model for the European power system},''
  {\em FME CenSES working paper series}, 2016.

\bibitem{Skar2014}
C.~Skar, G.~Doorman, and A.~Tomasgard, ``{The future European power system
  under a climate policy regime},'' in {\em {2014 IEEE International Energy
  Conference (ENERGYCON)}}, pp.~318--325, 2014.

\bibitem{projectGithub}
``{EMPIRE-Public}.'' \url{https://github.com/Goggien/EMPIRE-Public}.
\newblock Accessed 12.05.2022.

\bibitem{Birge2011}
J.~R. Birge and F.~Louveaux, {\em {In troduction to Stochastic Programming}}.
\newblock Springer Series in Operations Research and Financial Engineering, New
  York, NY: Springer New York, 2011.

\bibitem{Kaut2014b}
M.~Kaut, K.~T. Midthun, A.~S. Werner, A.~Tomasgard, L.~Hellemo, and M.~Fodstad,
  ``{Multi-horizon stochastic programming},'' {\em Computational Management
  Science}, vol.~11, no.~1, pp.~179--193, 2014.

\bibitem{Wang2021}
A.~Wang, J.~Jens, D.~Mavins, M.~Moultak, M.~Schimmel, K.~{Van Der Leun},
  D.~Peters, and M.~Buseman, ``{European Hydrogen Backbone: Analysing future
  demand, supply, and transport of hydrogen},'' Tech. Rep. June, Gas for
  Climate, Utrecht, 2021.

\bibitem{Buttler2018}
A.~Buttler and H.~Spliethoff, ``{Current status of water electrolysis for
  energy storage, grid balancing and sector coupling via power-to-gas and
  power-to-liquids: A review},'' {\em Renewable and Sustainable Energy
  Reviews}, vol.~82, pp.~2440--2454, February 2018.

\bibitem{LondonEconomics2013}
{London Economics}, ``{The Value of Lost Load (VoLL) for Electricity in Great
  Britain: Final report for OFGEM and DECC},'' Tech. Rep. July, London
  Economics, 2013.

\bibitem{Bertuccioli2014}
L.~Bertuccioli, A.~Chan, D.~Hart, F.~Lehner, B.~Madden, and E.~Standen,
  ``{Development of Water Electrolysis in the European Union},'' tech. rep.,
  E4tech {\&} Elemental Energy, 2014.

\bibitem{4COffshore}
{4C Offshore}, ``{Global Offshore Renewable Map}.''
  \url{https://map.4coffshore.com/offshorewind/}.
\newblock Accessed 12.12.2021.

\bibitem{Borrmann2018}
R.~Borrmann, K.~Rehfeldt, A.-K. Wallasch, and S.~Lüers, ``{Capacity Densities
  of European Offshore Wind Farms},'' tech. rep., Deutsche Windguard, 5 2018.

\bibitem{hartel2021}
P.~H{\"a}rtel, {\em {Offshore Grids in Low-Carbon Energy Systems. Long-Term
  Transmission Expansion Planning in Energy Systems with Cross-Sectoral
  Integration using Decomposition Algorithms and Aggregation Methods for
  Large-Scale Optimisation Problems.}}
\newblock PhD thesis, University of Kassel, April 2021.

\bibitem{Troncoso2011ElectrolysersHydrogen}
E.~Troncoso and M.~Newborough, ``{Electrolysers for mitigating wind curtailment
  and producing ‘green’ merchant hydrogen},'' {\em International Journal of
  Hydrogen Energy}, vol.~36, pp.~120--134, 1 2011.

\bibitem{Mathiesen2015SmartSolutions}
B.~V. Mathiesen, H.~Lund, D.~Connolly, H.~Wenzel, P.~A. Ostergaard,
  B.~M{\"{o}}ller, S.~Nielsen, I.~Ridjan, P.~Karn{\o}e, K.~Sperling, and F.~K.
  Hvelplund, ``{Smart Energy Systems for coherent 100{\%} renewable energy and
  transport solutions},'' {\em Applied Energy}, vol.~145, pp.~139--154, 5 2015.

\bibitem{Connolly2016SmartUnion}
D.~Connolly, H.~Lund, and B.~V. Mathiesen, ``{Smart Energy Europe: The
  technical and economic impact of one potential 100{\%} renewable energy
  scenario for the European Union},'' {\em Renewable and Sustainable Energy
  Reviews}, vol.~60, pp.~1634--1653, 7 2016.

\bibitem{Thellufsen2020SmartContext}
J.~Z. Thellufsen, H.~Lund, P.~Sorkn{\ae}s, P.~A. {\O}stergaard, M.~Chang,
  D.~Drysdale, S.~Nielsen, S.~R. Dj{\o}rup, and K.~Sperling, ``{Smart energy
  cities in a 100{\%} renewable energy context},'' {\em Renewable and
  Sustainable Energy Reviews}, vol.~129, p.~109922, 9 2020.

\bibitem{Sorkns2020SmartMarkets}
P.~Sorkn{\ae}s, H.~Lund, I.~R. Skov, S.~Dj{\o}rup, K.~Skytte, P.~E. Morthorst,
  and F.~Fausto, ``{Smart Energy Markets - Future electricity, gas and heating
  markets},'' {\em Renewable and Sustainable Energy Reviews}, vol.~119,
  p.~109655, 3 2020.

\end{thebibliography}

\end{document}